\pdfoutput=1
\documentclass[11pt,a4paper]{article}
\usepackage{jheppub,slashed}

\newcommand{\nn}{\nonumber}
\newcommand{\Mpl}{\overline{M}_{\rm Pl}}
\newcommand{\gld}{\tilde G}
\newcommand{\go}{\tilde g}
\newcommand{\sq}{\tilde q}
\newcommand{\Etmiss}{\slashed{E}_T}

\def\del{\partial}

\newcommand{\idttheta}{{\int d^2 \theta \,}}
\newcommand{\idftheta}{{\int d^4 \theta \,}}
\newcommand{\hc}{{\text{h.c.}}}

\preprint{
 CP3-15-03, MCnet-15-01}

\title{
Signals of a superlight gravitino at the LHC 
}

\author[a]{Fabio Maltoni,}
\emailAdd{fabio.maltoni@uclouvain.be} 
\author[a]{Antony Martini,} 
\emailAdd{antony.martini@uclouvain.be} 
\author[b]{Kentarou Mawatari}
\emailAdd{kentarou.mawatari@vub.ac.be}
\author[b]{and Bettina Oexl}
\emailAdd{bettina.oexl@vub.ac.be}

\affiliation[a]{Centre for Cosmology, 
 Particle Physics and Phenomenology (CP3),\\
 Universit\'e catholique de Louvain, 
 Chemin du Cyclotron 2, B-1348 Louvain-la-Neuve, Belgium}
\affiliation[b]{Theoretische Natuurkunde and IIHE/ELEM, 
 Vrije Universiteit Brussel,\\
 and International Solvay Institutes, Pleinlaan 2, 
 B-1050 Brussels, Belgium}

\abstract{
Very light gravitinos could be produced at a sizeable rate at colliders and have been searched
for in the mono-photon or mono-jet plus missing momentum signature. Strategies for enhancing the signal over background
and  interpretations of the experimental results are typically obtained within an effective field theory approach where 
all SUSY particles except the gravitino are heavy and are not produced resonantly.
We extend this approach to a simplified model that includes squarks and gluinos in the TeV range. 
In such a case, the jet(s)-plus-missing-momentum signature can be generated  through three different concurring mechanisms: 
gravitino-pair production with an extra jet, associated gravitino production 
with a squark or a gluino, or squark/gluino pair production with their
subsequent decay to a gravitino and a jet.  
By using a matrix-element parton-shower merging procedure, we take into account all the relevant production processes consistently, explore the SUSY parameter space with the LHC Run-I data set, and give prospects for the Run II. 
We also consider the reach of other signatures involving electroweak
particles, e.g., the mono-photon, -$Z$, or -$W$ plus missing momentum. The current mono-jet and mono-photon LHC analyses are interpreted to set a lower bound on the gravitino mass. 
We show how the limit of $m_{3/2}>1.7\times10^{-13}$~GeV obtained in the effective field theory hypothesis is modified  when the squarks and/or the gluino are in the TeV range. 
}

\begin{document}
\maketitle

\section{Introduction}\label{sec:intro}

Events with large missing momentum, and in particular those featuring
just one visible recoiling object (a jet, a photon, a weak boson, a top quark), are among 
the most promising final states where to look for signs of new physics at colliders.
Their simplicity and model independent nature appeal to both theorists and experimentalists. 
There are, however, important challenges that have to be faced with such signatures.
The first ones are of experimental nature. The accurate and precise
determination of the missing momentum
in events needs a detailed control of many aspects, from triggering to jet energy scales, to 
underlying event simulation, to  pile-up mitigation. In addition, in case of weak boson or top quark, 
tagging and reconstruction efficiencies for the recoiling object(s) also enter. 
The second class of challenges are more of theoretical nature and have to do with the problem of maximising the information
that can be extracted from data to constrain new physics models. 
Model-independent searches for dark matter (DM) constitute the most
popular interpretations of mono-jet analyses at the LHC, 
both in an effective field theory (EFT) framework as well as in simplified models, 
see e.g.~\cite{DiFranzo:2013vra,Abdallah:2014hon,Malik:2014ggr} and the references therein.  

Among new complete physics scenarios leading to mono-object plus large missing momentum
signals, supersymmetric (SUSY) models with a very light (or {\it
superlight})  gravitino play a special role: they offer a concrete
setting  where  the strengths and limitations  of EFT approach vis-a-vis
more UV completed models can be studied in detail.  

Let us look closer at model constraints from mono-object
searches at previous and current colliders.
At the LEP collider, the mono-photon signal was used to set a limit on models of
SUSY with the gravitino as the lightest SUSY particle (LSP) and extra
dimensions~\cite{Abbiendi:2000hh,Heister:2002ut,Achard:2003tx,Abdallah:2003np}. 
In some SUSY scenarios the gravitino can be very light
of order $m_{3/2}\sim{\cal O}(10^{-14}-10^{-12})$~GeV with all the other SUSY particles being
above the TeV threshold~\cite{Nachtmann:1984xu,Brignole:1997sk}. We dub such a scenario as {\it ``gravitino EFT''}.
The only relevant parameter in this case is the gravitino mass, which is directly related to the SUSY breaking scale, 
the lower limit being $m_{3/2}>1.35\times10^{-14}$~GeV~\cite{Achard:2003tx,Agashe:2014kda}.
Alternatively, we consider the gravitino LSP in 
the minimal supersymmetric standard model
(MSSM) with other sparticles at the TeV scale.
In this scenario, the process of the neutralino--gravitino associated production with the
subsequent neutralino decay into a photon and a gravitino has been used to put a
limit on the gravitino mass as a function of the neutralino and selectron
masses~\cite{Fayet:1986zc,Dicus:1990vm,Lopez:1996gd,Lopez:1996ey,Baek:2002np,Mawatari:2011cu}, e.g. 
$m_{3/2}\gtrsim 10^{-14}$~GeV
for $m_{\tilde\chi^0_1}=140$~GeV and $m_{\tilde e}=150$~GeV~\cite{Abdallah:2003np}.
Such a scenario can also be considered  as a simplified SUSY model, where only the
gravitinos, the lightest neutralino and the selectrons play a role in the phenomenology at colliders.

At the Tevatron, not only the mono-photon but also the mono-jet signals
constrain models of SUSY~\cite{Affolder:2000ef,Acosta:2002eq} and extra
dimensions~\cite{Acosta:2002eq,Abazov:2008kp,Aaltonen:2008hh}.
Similar to the LEP bound, in the gravitino-EFT limit~\cite{Brignole:1998me} a gravitino is excluded below
$1.1\times10^{-14}$~GeV and  $1.17\times10^{-14}$~GeV in the
mono-jet~\cite{Affolder:2000ef} and mono-photon~\cite{Acosta:2002eq} channels, respectively.

At the LHC, besides the mono-photon~\cite{Khachatryan:2014rwa,Aad:2014tda} and 
mono-jet~\cite{ATLAS:2012zim,Khachatryan:2014rra} signals, other
mono-object plus missing transverse momentum signals such as a $Z$
boson~\cite{Aad:2014vka}, a lepton~\cite{ATLAS:2014wra,Khachatryan:2014tva}, and
a top quark~\cite{Khachatryan:2014uma} have been investigated mostly in the context 
of DM searches and more exotic models.  SUSY models have been considered only in the ATLAS mono-jet
analysis~\cite{ATLAS:2012zim}, where the gluino--gravitino~\cite{Dicus:1989gg,Drees:1990vj,Dicus:1996ua,Kim:1997iwa,Klasen:2006kb,Mawatari:2011jy} and squark--gravitino~\cite{Kim:1997iwa,Klasen:2006kb} associated productions were
taken into account to set a limit on the gravitino mass as a function of the squark and gluino masses as, e.g. 
\begin{align}
 m_{3/2}> 1\times10^{-13}\ (4\times 10^{-14})\ {\rm GeV}
\end{align}
for the degenerate squark and gluino masses at $m_{\sq,\go}=500$ (1700)~GeV.
One point that is relevant for this work is  that the above limit may be modified
by the contribution from the direct gravitino-pair
production in association with an extra jet, a production channel so far disregarded in the analysis. 
Moreover, while event selection is targeted to the associated
gravitino production, events from squark and gluino pair production
may enter the signal region affecting the results.  

We would like to put forward the interpretation of the
mono-object signals in the SUSY context with a very light gravitino
for the LHC.
As mentioned above, extending the gravitino EFT to the full MSSM 
(or simplified SUSY models), other production channels can contribute
leading to rather different  final state features that in turn depend on the SUSY parameters.  
In Ref.~\cite{deAquino:2012ru} we studied the gluino--gravitino and gluino-pair
production in this very same context.  However, the gravitino-pair
production associated with a jet and 
squark--gravitino production were not included there. 
In this work we present, for the first time, the complete set of production channels consistently treated
in a unique framework that can provide accurate predictions for the
general case. In addition, although the gravitino in our scenario is too
light to be a cold DM candidate, the approach we have followed is fully general and 
can be used as a template for passing from an EFT approach to simplified models in the context of DM searches~\cite{Papucci:2014iwa}.

The plan of the paper is as follows. In sec.~\ref{sec:theory} we  focus on the SUSY QCD sector in order to
assess the parameter space relevant  for gravitino production processes and 
potentially contributing to the mono-jet signature. We explicitly construct a SUSY QCD model in sec.~\ref{sec:model}.
In sec.~\ref{sec:production} we present the three different yet related mechanisms which
produce gravitinos.  Gravitino-pair production with one jet has been studied in the gravitino-EFT limit only~\cite{Brignole:1998me}, where exact tree-level results for $2\to3$ matrix elements for 
$p\bar p/pp\to\gld\gld j$ have been computed only for the quark--antiquark and quark--gluon initial states, but not for gluon--gluon 
ones. We obtain such results for generic squark/gluino masses and for
all processes for the first time in this work.
In sec.~\ref{sec:tool} we briefly review the computation/simulation tools used in this article.
In sec.~\ref{sec:jet} we study all the relevant gravitino
production processes in detail for  total as well as differential cross sections. As an application of our results, we 
recast the ATLAS mono-jet analysis~\cite{ATLAS:2012zim} with inclusive signal samples by merging
matrix elements with parton showers (ME+PS) in order to set a limit on the masses of the SUSY particles. 
We suggest improvements to the analysis so to increase the sensitivity
to the gravitino mass when squarks and gluinos are light.
In sec.~\ref{sec:photon}, we consider the associated production with an electroweak (EW) particle, and
study the mono-photon, -$Z$ and -$W$ signals in the very light
gravitino context.  Finally, we recast the mono-photon analyses at the
LHC~\cite{Khachatryan:2014rwa,Aad:2014tda} to set a limit on the gravitino
mass. Section~\ref{sec:summary} is devoted to our conclusions.

\section{Light gravitino production at the LHC}
\label{sec:theory}

In this section we start by constructing a SUSY QCD model by using the superspace formalism.
We then present the three mechanisms of light gravitino production at hadron colliders and finally
we briefly describe the simulation tools we employ for our results. 

\subsection{SUSY QCD with a goldstino superfield}
\label{sec:model}

In phenomenologically viable SUSY models, SUSY breaking is often
assumed to take place in a so-called hidden sector, and then 
transmitted to the visible sector (i.e. the SM particles and their
superpartners) through some mediation mechanism, e.g. gauge mediation or
gravity mediation. As a result, one
obtains effective couplings of the fields in the visible sector to the
goldstino multiplet. 
To illustrate the interactions among the physical
degrees of freedom of the goldstino multiplet and the fields in the
visible sector, we introduce an $R$-parity conserving $N=1$ global
supersymmetric model with the $SU(3)_{C}$ gauge group in the
superspace formalism. The model comprises one vector superfield
$V=(A^{\mu},\lambda,D_V)$, describing a gluon $A^{\mu}$ and a gluino
$\lambda$, and two chiral superfields $\Phi_{L}=(\sq_{L},q_{L}, F_{L})$
and $\Phi_{R}=(\sq^*_{R},q_{R}^{c},F_{R})$, containing the left- and 
right-handed quarks $q_{L/R}$ and squarks $\sq_{L/R}$, where the color
and generation indices are suppressed. In
addition, we introduce a chiral superfield in the hidden sector
$X=(\phi,\gld,F_X)$, containing a sgoldstino $\phi$ and a goldstino
$\gld$. $D_V$, $F_{L/R}$ and $F_X$ are auxiliary fields.

The Lagrangian of the visible sector is 
\begin{align}
 \mathcal{L}_{\rm vis}=
 \sum_{i=L,R}\idftheta\,\Phi^{\dagger}_ie^{2g_sV}\Phi_i 
 +\Big(\frac{1}{16g_s^2}\idttheta\,W^{\alpha} W_{\alpha} +\hc\Big),
\label{L_vis}
\end{align}
where $g_s$ is the strong coupling constant.%
\footnote{The covariant derivative is defined as
$D_{\mu}=\del_{\mu}+ig_sT^aA_{\mu}^a$.}
$W_{\alpha}=-\frac{1}{4}\bar{D}\cdot\bar{D}\,e^{-2g_sV}D_{\alpha}\,e^{2g_sV}$ denotes the SUSY
$SU(3)_{C}$ field strength tensor with $D$ being the
superderivative. $\mathcal{L}_{\rm vis}$ contains the kinetic terms as
well as the gauge interactions. 

The Lagrangian of the goldstino part is given by
\begin{align}
 \mathcal{L}_{X}=
 \idftheta\,X^{\dagger}X-\Big(F\idttheta\,X+\hc\Big) 
 -\frac{c_X}{4}\idftheta\,(X^{\dagger}X)^2.
\label{L_hid}
\end{align}
The first term gives the kinetic term of the sgoldstino and the goldstino, while
the second term is a source of SUSY breaking and 
$F\equiv\langle F_X\rangle$ is a vacuum expectation value (VEV) of
$F_X$.%
\footnote{Note that we follow the {\sc FeynRules} convention for chiral 
superfields 
$\Phi(y,\theta)=\phi(y)+\sqrt{2}\,\theta\cdot\psi(y)-\theta\cdot\theta\,F(y)$~\cite{Alloul:2013bka}, 
which fixes the sign of the Lagrangian so as to give a positive
contribution to the scalar potential.} 
The last term is non-renormalizable and provides interactions in
the goldstino multiplet. In addition, this term also gives the sgoldstino mass term
when replacing the auxiliary fields $F_X$ by the VEV, and hence we
assign $c_X=m^2_{\phi}/F^2$. 

The effective Lagrangian that leads to
the interactions among the (s)goldstinos and the fields in the visible
sector as well as the soft mass terms for the squarks and the gluinos
is given by 
\begin{align}
 \mathcal{L}_{\text{int}}=-\sum_{i=L,R}c_{\Phi_i}\idftheta\, 
  X^{\dagger}X\Phi_i^{\dagger}\Phi_i
 -\Big(\frac{c_V}{16g_s^2}\idttheta\, X W^{\alpha} W_{\alpha} +\hc\Big),
\label{L_int}
\end{align}
where we identify $c_{\Phi_i}=m^2_{\sq_i}/F^2$ and $c_V=2m_{\lambda}/F$. 
We note that our model is minimal, yet enough to generate all the
relevant interactions involving two goldstinos in the final state 
for the jet(s)$+\Etmiss$ signal at hadron colliders.
The extension of the model including the SM electroweak (EW) gauge
group is straightforward, and we will study mono-$\gamma$, -$W$ and -$Z$
signals later.

Let us briefly refer to the
goldstino equivalence theorem. When the global SUSY is promoted to the
local one, the goldstino is absorbed by the gravitino via the so-called
super-Higgs mechanism. 
In the high-energy limit, $\sqrt{s}\gg m_{3/2}$, 
the interactions of
the helicity 1/2 components are dominant, and can be well described by
the goldstino interactions due to the gravitino-goldstino equivalence
theorem~\cite{Casalbuoni:1988kv,Casalbuoni:1988qd}. 
As a
consequence of the super-Higgs mechanism, the gravitino mass is related
to the SUSY breaking scale and the Planck mass 
as~\cite{Volkov:1973jd,Deser:1977uq}
\begin{align}
 m_{3/2}=\frac{F}{\sqrt{3}\,\Mpl},
\label{grav_mass}
\end{align}
where $\Mpl\equiv M_{\rm Pl}/\sqrt{8\pi}\approx 2.4\times10^{18}$~GeV is
the reduced Planck mass. Therefore, low-scale SUSY breaking scenarios
such as GMSB
provide a gravitino LSP. In the following, we simply call the goldstino
``gravitino''.

\subsection{Light gravitino production}
\label{sec:production}

\begin{figure}
\center
 \includegraphics[width=.7\textwidth,clip]{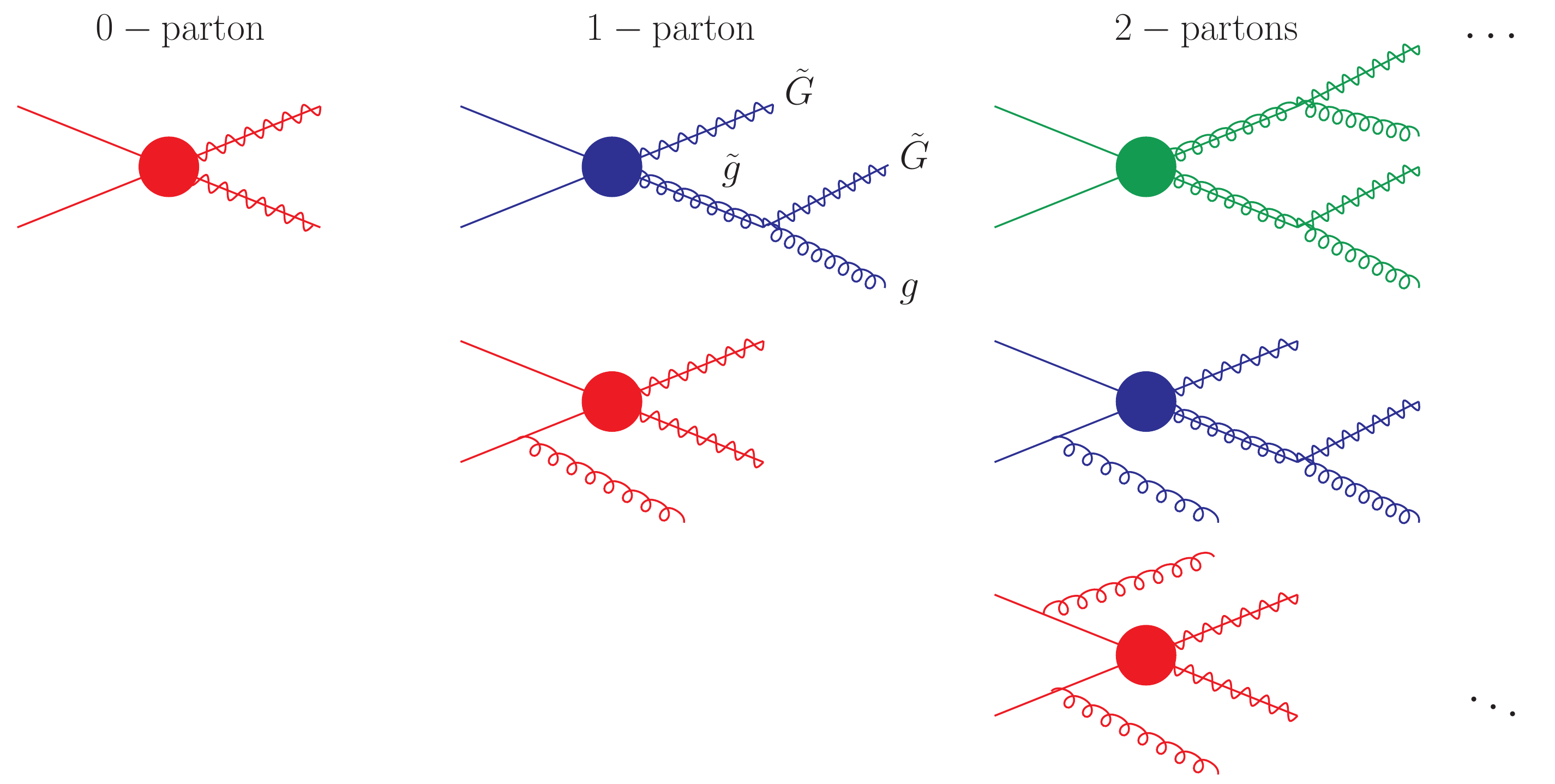} 
 \caption{Schematic diagrams for $pp\to\gld\gld+0,1,2$ partons. In the
 first row the leading gravitino-pair (red),
 gluino--gravitino (blue) and gluino-pair (green)
 diagrams are sorted. The diagrams are ordered with the number of
 additional QCD partons in rows, while with the total parton
 multiplicity in columns.
}
\label{fig:diagram}
\end{figure}

Given the model we constructed in the previous section,
we now consider light-gravitino production in $R$-parity conserving scenarios
that lead to jet(s) plus  missing momentum at the LHC: 
\begin{align}
 pp\to {\rm jet(s)}+\Etmiss,
\end{align}
where the missing momentum is carried by two LSP gravitinos.
At the leading order in QCD, the relevant processes are:
\begin{enumerate}
\item gravitino-pair production in association with a quark/gluon emission from initial state radiation,
\item gravitino production associated with a squark/gluino with the
      subsequent decay into  a gravitino and a quark/gluon, 
\item SUSY QCD pair production with the subsequent decay into gravitino and a quark/gluon. 
\end{enumerate}
The processes are schematically represented in fig.~\ref{fig:diagram}. 
The processes in the second column of fig.~\ref{fig:diagram}
contribute in an obvious way to the mono-jet signal. However, also the 2-parton final states will contribute either in the exclusive
1-jet analysis because one parton might not give rise to a jet or when the analysis is fully or in part inclusive over other jets.
In the current ATLAS and CMS  mono-jet analyses~\cite{ATLAS:2012zim,Khachatryan:2014rra}, for example, a second jet is allowed and hence those events potentially fall into the signal region. For the mono-$\gamma$, -$Z$ and -$W$ signals, we simply replace the QCD processes by the EW processes, i.e. replace gluinos by neutralinos and charginos.  

We now consider each production channel in more detail.

\subsubsection{Gravitino pair production}
\label{sec:gldgld}

Direct gravitino-pair production at colliders has been studied only in
models where all SUSY particles except for the gravitino are too heavy
to be produced on-shell, i.e. in the gravitino EFT limit~\cite{Nachtmann:1984xu,Brignole:1997sk,Brignole:1998me}. 
One of the aims of this article is to extend the previous studies to
take into account the effect of other SUSY particles in spectrum. 
This has been done recently for mono-photon signals at future linear colliders~\cite{Mawatari:2014cja}, 
and we now apply for it to the QCD sector for the LHC. 

A pair of gravitinos is produced through both the $q\bar q$ and $gg$
initial states,
\begin{align}
 pp(q\bar q, gg)\to\gld\gld,
\label{process_gldgld}
\end{align}
and can be observed if extra radiation is hard enough to be detected, for instance  in the form of one or more jets. 
The helicity amplitudes for the above $2\to2$ processes were presented in terms of the $e^+e^-$ and
$\gamma\gamma$ initial states in~\cite{Mawatari:2014cja}. A remarkable feature of this production channel 
is that the corresponding total cross section scales as the inverse of the gravitino mass to the fourth power,
\begin{align}
 \sigma(\gld\gld)\propto 1/m_{3/2}^4.
\label{xsec_gldgld}
\end{align}
Another feature is that the cross section tends to be larger for heavier
squarks and gluinos, which are propagating in the $t$ and $u$ channels. 
For the $gg$ channel, there are diagrams featuring $s$-channel sgoldstino.  These
play an important role in the computation of cross sections even when sgoldstinos are too
heavy to be produced; see ref.~\cite{Mawatari:2014cja} for more details.

As expected from the colourless nature of the gravitinos, an extra parton in the final state
mainly comes from initial state radiation and therefore it is naturally suppressed by $\alpha_S/p_T^4$.  
Hence the mono-jet rate from this process strongly depends on the jet minimum $p_T$ (or equivalently
from the minimum missing momentum).  We will investigate those effects carefully in sec.~\ref{sec:jet}. 
The $2\to3$ processes
\begin{align}
 pp(q\bar q, qg, gg)\to\gld\gld j,
\label{process_gldgldj}
\end{align}
have been calculated for $q\bar q$ and $qg$ initial states 
in the gravitino EFT limit, yet the $gg$
process was estimated only by the $2\to2$ cross section in the limit of the
soft and collinear gluon radiation~\cite{Brignole:1998me}.
In this article, as shown later, we consider all the amplitudes at tree-level 
without any approximation  and calculate the full matrix elements numerically.

\subsubsection{Associated gravitino production}

Gravitino production in association with a squark or a gluino and the
subsequent decay into a gravitino and a quark/gluon, 
\begin{align}
 pp\to\sq\gld,\go\gld\to\gld\gld j,
\label{process_gogld}
\end{align}
leads to the $j+\Etmiss$ signal at the leading order (LO), and has been
studied in~\cite{Dicus:1989gg,Drees:1990vj,Dicus:1996ua,Kim:1997iwa,Klasen:2006kb,Mawatari:2011jy}.
The tree-level ME+PS merging technique has also been applied for this
process in~\cite{deAquino:2012ru}.

Unlike the gravitino-pair production in eq.~\eqref{xsec_gldgld}, the
cross section is inversely proportional to the square of the gravitino
mass, 
\begin{align}
 \sigma(\sq\gld,\go\gld)\propto 1/m_{3/2}^2,
\label{xsec_gogld}
\end{align}
and hence the dependence of the gravitino mass is milder than in the
gravitino-pair production.  Similar to the $\gld\gld$ production, heavier squarks and gluinos in the
$t$ and $u$ channels enhance the cross sections, while those in the
final state suppress the cross sections due to the phase space.

\subsubsection{Indirect gravitino production}

SUSY QCD pair productions, i.e. squark-pair, gluino-pair and
squark--gluino productions, have  been systematically studied, motivated by the
inclusive SUSY searches as well as in simplified SUSY searches. On the other hand, they
have not been considered in the mono-jet analysis since more than one jet in the final
state is expected. Especially, when squarks and/or gluinos are the next-to-lightest SUSY
particle (NLSP), their decay can provide
the di-jet plus missing momentum signal at the LO~\cite{Baer:1998pg,Kats:2011qh}:
\begin{align}
 pp\to\sq\sq,\sq\go,\go\go\to\gld\gld jj.
\label{process_gogo}
\end{align} 
As mentioned above, in the current mono-jet analyses by
ATLAS~\cite{ATLAS:2012zim} and CMS~\cite{Khachatryan:2014rra}, events
with a second jet have been included as the signal typically contains
more jets from QCD radiation. Therefore, depending on cuts, the jets coming from
the decay of heavy SUSY particles may contribute to the signal region. 
We also note that, when the gravitino is very light, the $t$-channel
gravitino exchange enhances the cross sections~\cite{Dicus:1989gg,Drees:1990vj,Dicus:1996ua,Kim:1997iwa}. 

\begin{figure}
\center
 \includegraphics[width=.5\textwidth,clip]{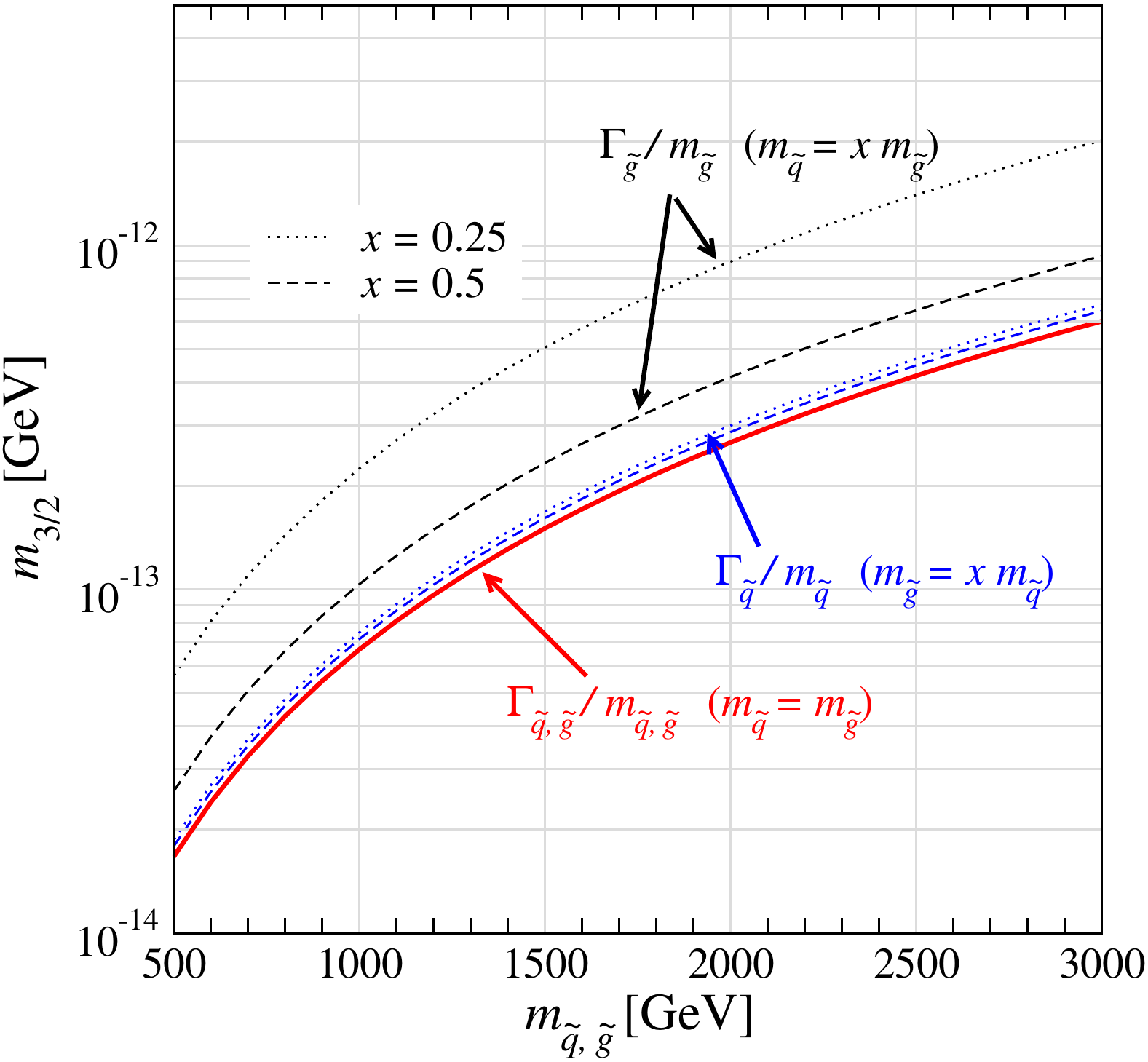} 
 \caption{25\% lines of the total width over the mass in the squark (gluino)
 and gravitino mass plane for
 $m_{\sq}=m_{\go}$ (red), $m_{\sq}>m_{\go}$ (blue), and $m_{\sq}<m_{\go}$
 (black) cases, where the main decay mode is
 $\sq(\go)\to q(g)+\gld$. 
 For the non-degenerate case, we take 
 $m_{\go}=\{4,\,2,\,1/2,\,1/4\}\times m_{\sq}$. 
 }
\label{fig:width}
\end{figure}

Before turning to the collider phenomenology part,
it may be worth to mention the decay width of the squark and gluino. The partial
decay width of a squark (gluino) into a quark (gluon) and a gravitino is
given by
\begin{align}
 \Gamma(\sq(\go)\to q(g)+\gld)=\frac{m^5_{\sq(\go)}}{48\pi\Mpl^2m^2_{3/2}},
\end{align}
where the gravitino mass in the phase space is neglected. When the gravitino is very light and/or the squarks and gluinos are
heavy, the width of the squark and gluino can be a significant fraction of the mass.
At the same time, the gravitino couplings become strong and the
perturbative calculations are not reliable. To identify a reasonable SUSY parameter space, 
in fig.~\ref{fig:width} we show $\Gamma/m=0.25$ lines for
$m_{\sq}=m_{\go}$ (red), $m_{\sq}>m_{\go}$ (blue), and $m_{\sq}<m_{\go}$
(black).%
\footnote{The widths are obtained numerically by the decay
package {\sc MadWidth}~\cite{Alwall:2014bza}.} 
We assume all other SUSY particles are heavier than the squarks and the
gluino. For the $m_{\sq}>m_{\go}$ case, the additional $\sq\to\go+q$ decay channel is
opened. 
For the $m_{\sq}<m_{\go}$ case, on the other hand, the gluino has
all possible squark decay modes, and hence its width becomes
significantly larger than the squark one,  strongly depending  on the
mass difference.
In the following, a benchmark scenario will be identified ($m_{3/2}=2\times10^{-13}$~GeV
with $m_{\sq}=m_{\go}=1$~TeV) where the widths are 28~GeV.

\subsection{Event simulation tools}
\label{sec:tool}

Here, we briefly describe event simulation tools we employ in this
article.  We follow the strategy presented in ref.~\cite{Christensen:2009jx} to new
physics  simulations.

Similar to the SUSY QED model of ref.~\cite{Mawatari:2014cja}, 
we have implemented the SUSY QCD Lagrangian with a goldstino supermultiplet
described in sec.~\ref{sec:model} into
{\sc FeynRules2}~\cite{Alloul:2013bka}, which provides the Feynman
rules in terms of the physical component fields and the {\sc UFO} model
file~\cite{Degrande:2011ua,deAquino:2011ub} for matrix-element
generators such as {\sc MadGraph5\_aMC@NLO}~\cite{Alwall:2014hca}.
In this work, instead of employing a dedicated implementation of the four-fermion vertices
involving more than one Majorana particle~\cite{Mawatari:2014cja}, we
introduce auxiliary heavy particles for the multi-jet simulation. 
Parton-level events generated by {\sc MadGraph5\_aMC@NLO} are passed to
{\sc Pythia6.4}~\cite{Sjostrand:2006za} for parton shower and hadronisation, to {\sc Delphes3}~\cite{deFavereau:2013fsa} for
detector simulation, and to {\sc MadAnalysis5}~\cite{Conte:2012fm} for sample analyses.

\section{Mono-jet plus missing momentum}\label{sec:jet}

In this section, we first present total and differential cross sections to
illustrate how the three gravitino production processes depend on the
SUSY mass parameters. Then, we recast the ATLAS mono-jet analysis~\cite{ATLAS:2012zim} to
constrain the gravitino mass in cases that go beyond the gravitino-EFT scenario.  

In the following, we consider three scenarios where squark and/or gluino masses are
${\cal O}(10)$~TeV and ${\cal O}(1)$~TeV:
\begin{subequations}
\begin{align}
 &{\rm A}:\quad m_{\sq}=m_{\go}=20~{\rm TeV}\qquad 
  &&\hspace*{-2.5cm}\text{(the gravitino-EFT limit)}, \\  
 &{\rm B}:\quad m_{\sq}=20~{\rm TeV},\ m_{\go}=1~{\rm TeV}\qquad
  &&\hspace*{-2.5cm}\text{(the heavy-squark limit)}, \\  
 &{\rm C}:\quad m_{\sq}=m_{\go}=1~{\rm TeV}, &  
\end{align}\label{cases}%
\end{subequations}
while we keep the sgoldstino masses at 20~TeV.
For simplicity, we assume that all the non-colored SUSY particles are heavier
than the colored ones, and hence the decay mode of squarks and gluinos is only
into gravitinos.
Only the gravitino-pair production contributes to the signal for case A, while the
$\go\gld$ and $\go\go$ productions can also give the signal for case B. 
In case C all the subprocesses can be comparable.
We note that the masses $m_{\sq}=m_{\go}=20$~TeV reproduce the results
of the total and differential cross sections 
in ref.~\cite{Brignole:1998me}, where all the SUSY particles except
gravitinos are integrated out, i.e. where the computation has been done in the
gravitino-EFT limit.

\subsection{Total rates}

\begin{figure}
\center
 \includegraphics[width=.495\textwidth,clip]{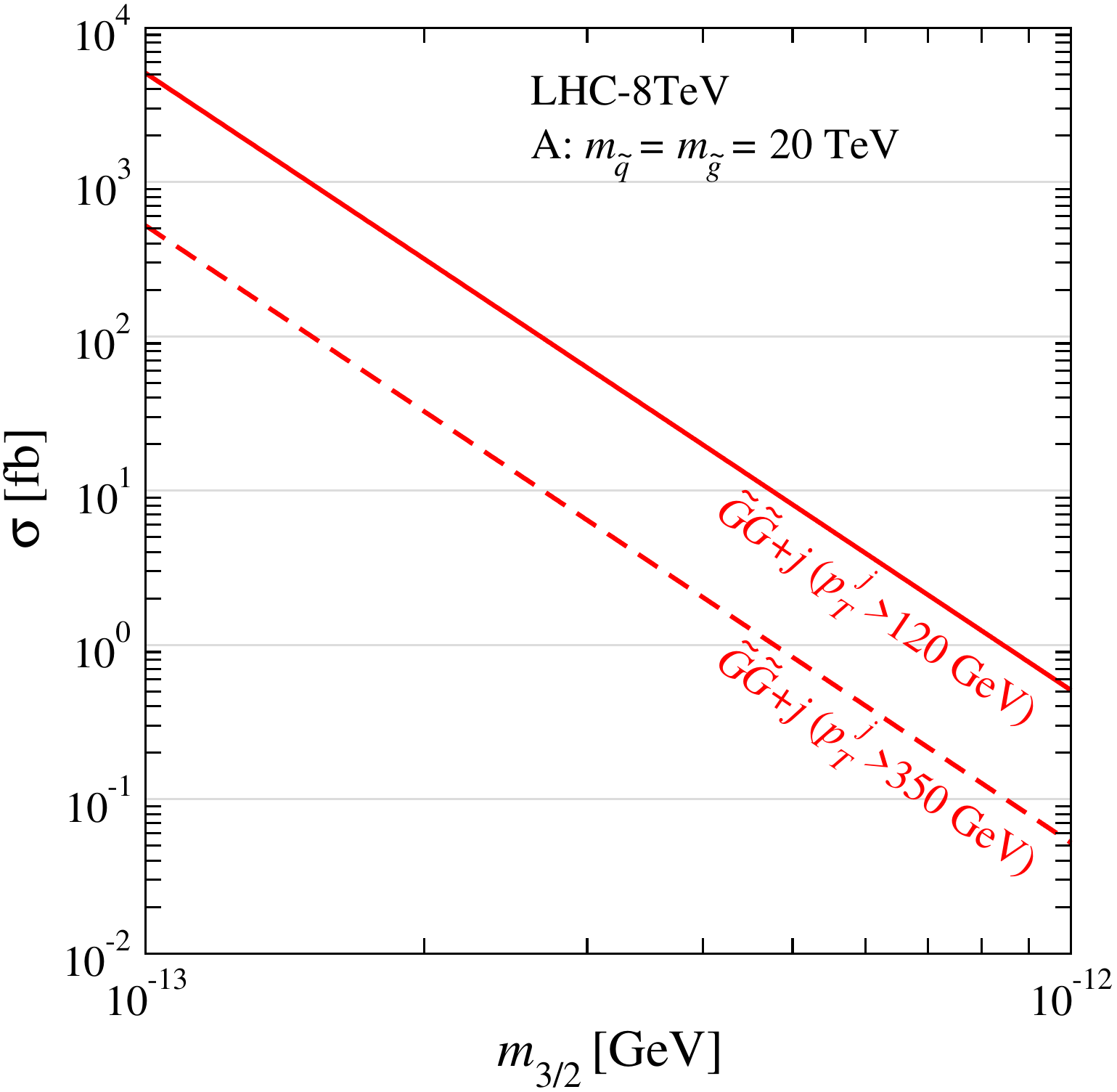} 
 \includegraphics[width=.495\textwidth,clip]{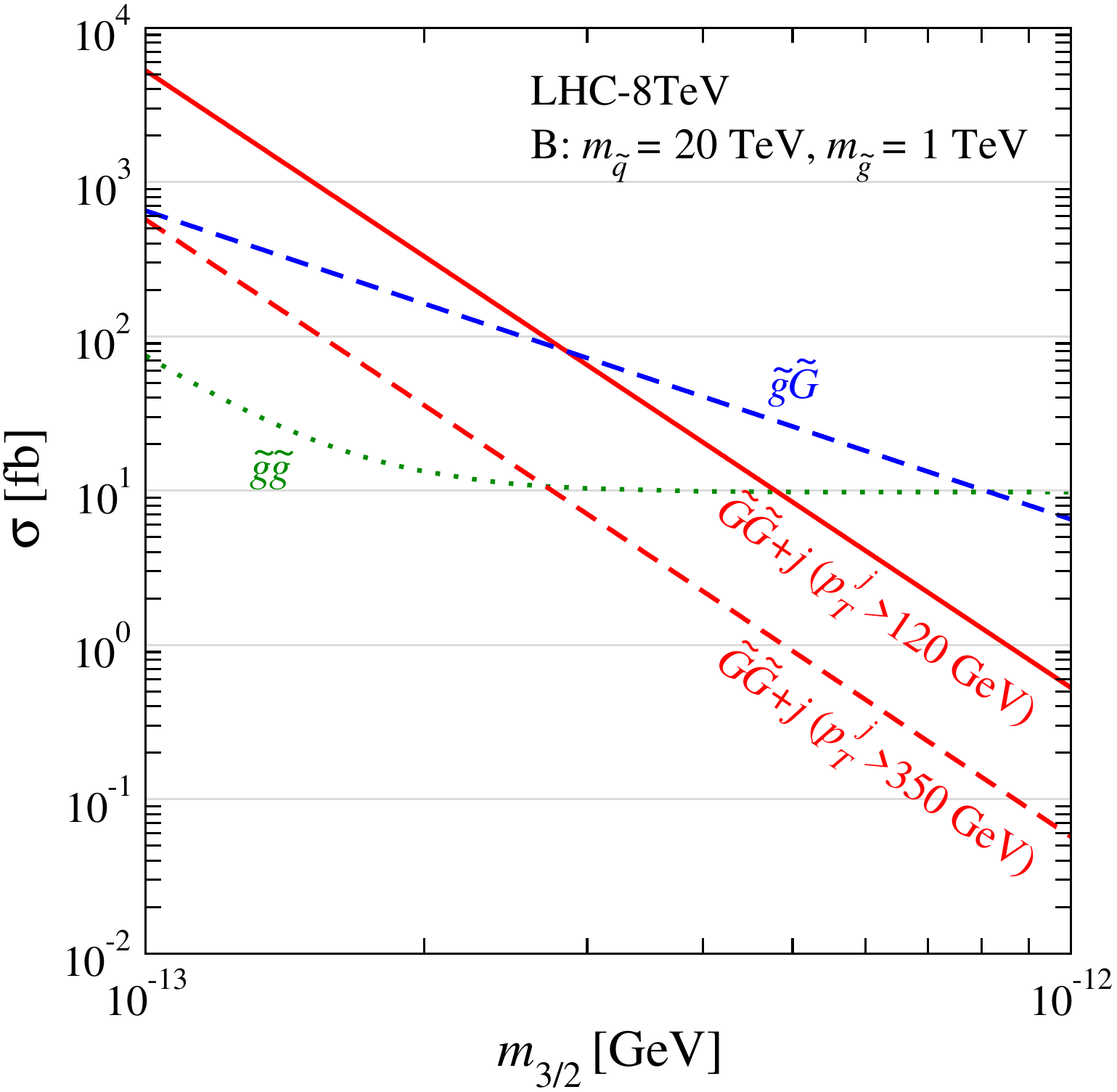} 
 \caption{Total cross sections of
 the gravitino-pair production with a QCD radiation ($\gld\gld+j$), 
 the gluino--gravitino associated production ($\go\gld$), and
 the gluino-pair production ($\go\go$) at $\sqrt{s}=8$~TeV 
 as a function of the gravitino mass
 for case A (left) and B (right).
 For the gravitino-pair production kinematical cuts
 $p_T^j>120/350$~GeV (solid/dashed) and $|\eta^j|<4.5$ are applied.
 }
\label{fig:xsec_mgld}
\end{figure}

\begin{figure}
\center
 \includegraphics[width=.495\textwidth,clip]{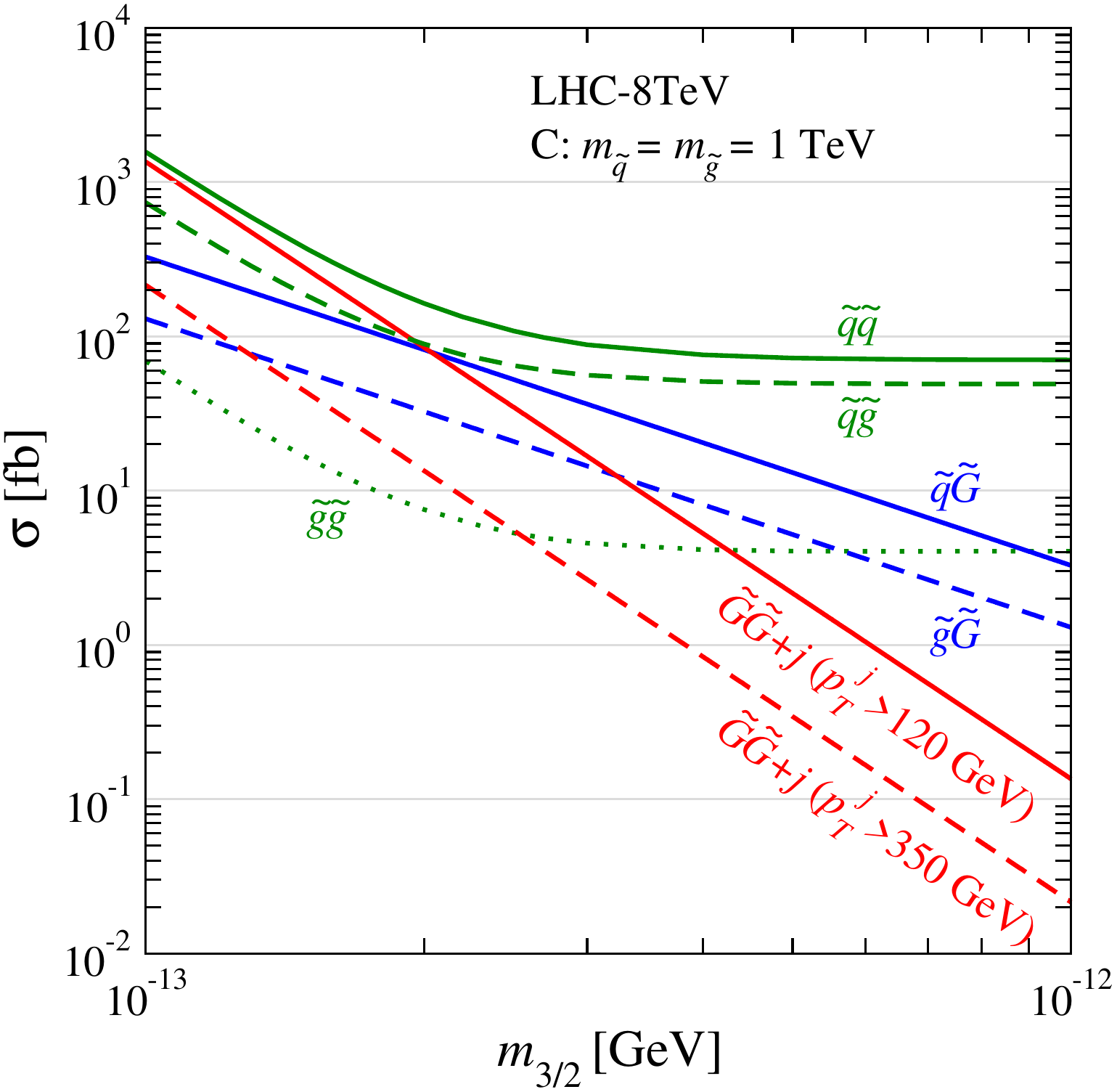} 
 \includegraphics[width=.495\textwidth,clip]{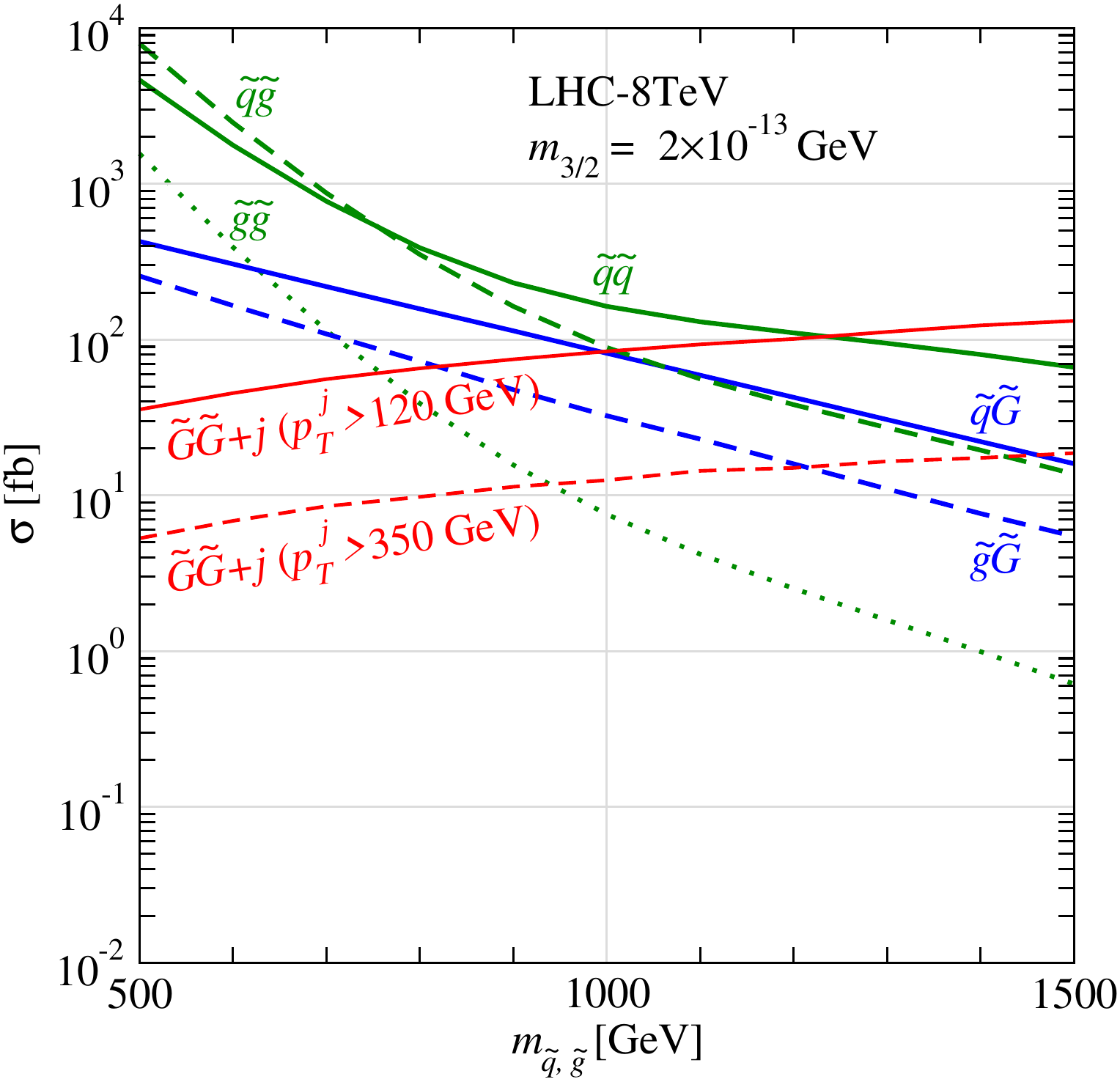} 
 \caption{Left: Same as fig.~\ref{fig:xsec_mgld}, but for case C.
 Right: Total cross sections as a function of the degenerate squark
 and gluino masses with the gravitino mass at  
 $m_{3/2}=2\times10^{-13}$~GeV. 
 }
\label{fig:xsec_msqgo}
\end{figure}

Figures~\ref{fig:xsec_mgld} and \ref{fig:xsec_msqgo} (left) show the total cross sections as a function
of the gravitino mass for the three scenarios at $\sqrt{s}=8$~TeV. 
For the gravitino-pair production plus an extra QCD emission
($\gld\gld+j$), we
impose a minimal transverse momentum cut for the jet with
$p_T^j>120$~GeV and 350~GeV in the region $|\eta^j|<4.5$.
We employ the CTEQ6L1 PDFs~\cite{Pumplin:2002vw} with the factorization and renormalization
scales at $p_T^j$ for the gravitino-pair production, 
$(m_{\sq,\go}+m_{3/2})/2\sim m_{\sq,\go}/2$ for the associated gravitino
production, and $(m_{\sq,\go}+m_{\sq,\go})/2\sim m_{\sq,\go}$ for the
SUSY QCD pair production. 
We note that all our results are the LO predictions although it is well
known that higher-order QCD corrections are large. 
For example, the $K$ factor of the gluino-pair production is about three
for $m_{\go}\sim 1$~TeV at the 8-TeV
LHC~\cite{Beenakker:1996ch,GoncalvesNetto:2012yt}, while the
higher-order calculations have not yet been done for the gravitino-pair 
production and the associated gravitino production.
Our analyses can be redone with different overall normalizations
and yet the main features will not change.

One can clearly see the $m_{3/2}^{-4}$ and $m_{3/2}^{-2}$ dependence for
the $\gld\gld(+j)$ and $\sq\gld/\go\gld$ processes, respectively,
as discussed in sec.~\ref{sec:production}. For the SUSY QCD pair
productions, $\sq\sq/\sq\go/\go\go$, the contribution of the $t$-channel
gravitino exchange can be visible if the 
gravitino is lighter than about $3\times10^{-13}$~GeV.

We also show the total rates as a function of the degenerate squark and
gluino masses with the fixed gravitino mass at $2\times10^{-13}$~GeV in
fig.~\ref{fig:xsec_msqgo} (right). 
For the gravitino-pair production, the cross section increases as the
squarks and 
gluinos become heavier.
On the other hand, the cross sections for the associated production and
the SUSY QCD pair production decreases due to the phase space
suppression. 

As can be seen in figs.~\ref{fig:xsec_mgld} and \ref{fig:xsec_msqgo}, 
each contribution to the total rates strongly depends on the SUSY mass
parameters, and 
the different contributions can be comparable for certain
parameters. 
However, the resulting signature can be still distinctive among the
subprocesses as shown below.

\subsection{Differential distributions}

\begin{figure}
\center
 \includegraphics[width=.328\textwidth,clip]{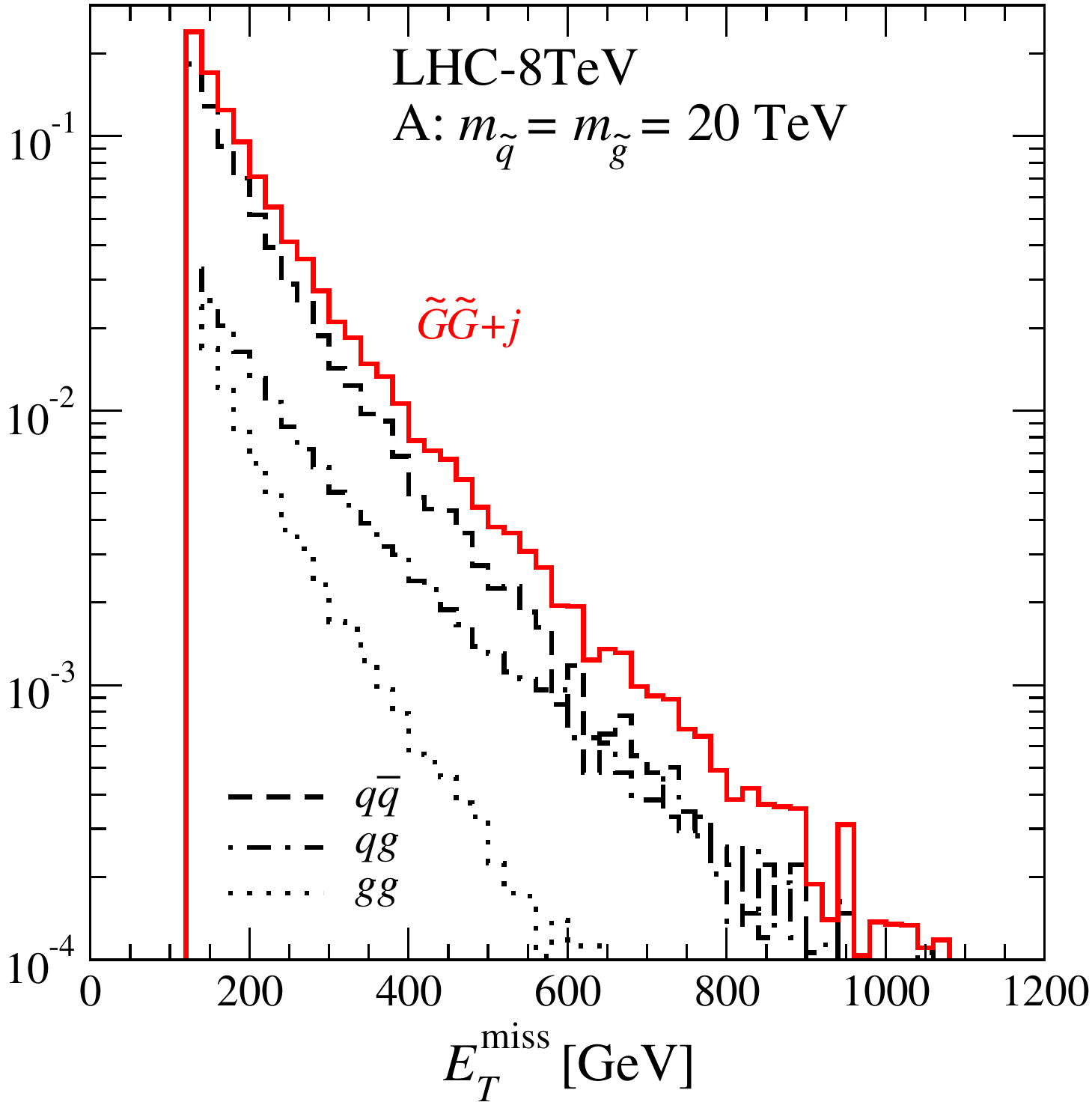}
 \includegraphics[width=.328\textwidth,clip]{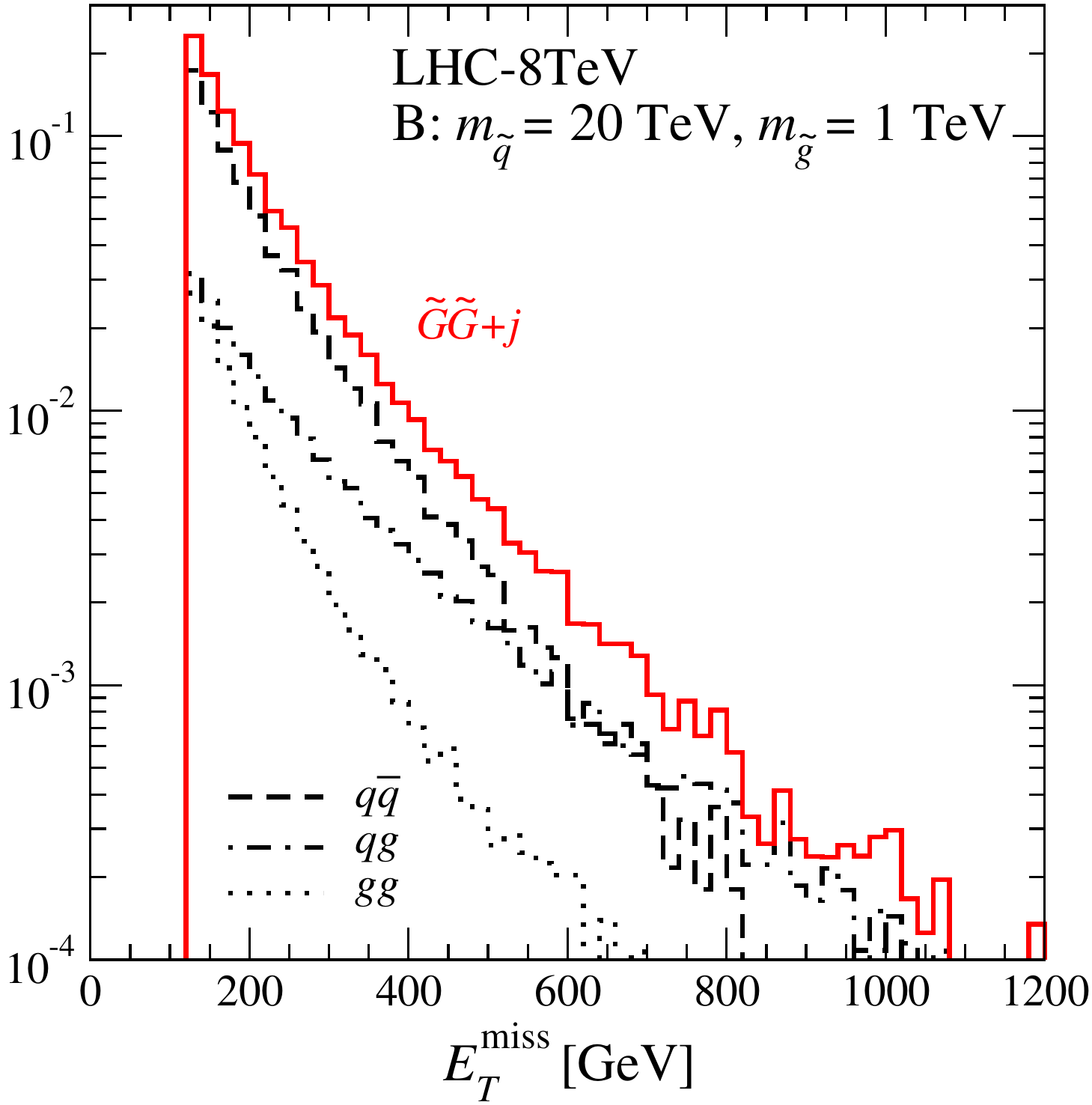}
 \includegraphics[width=.328\textwidth,clip]{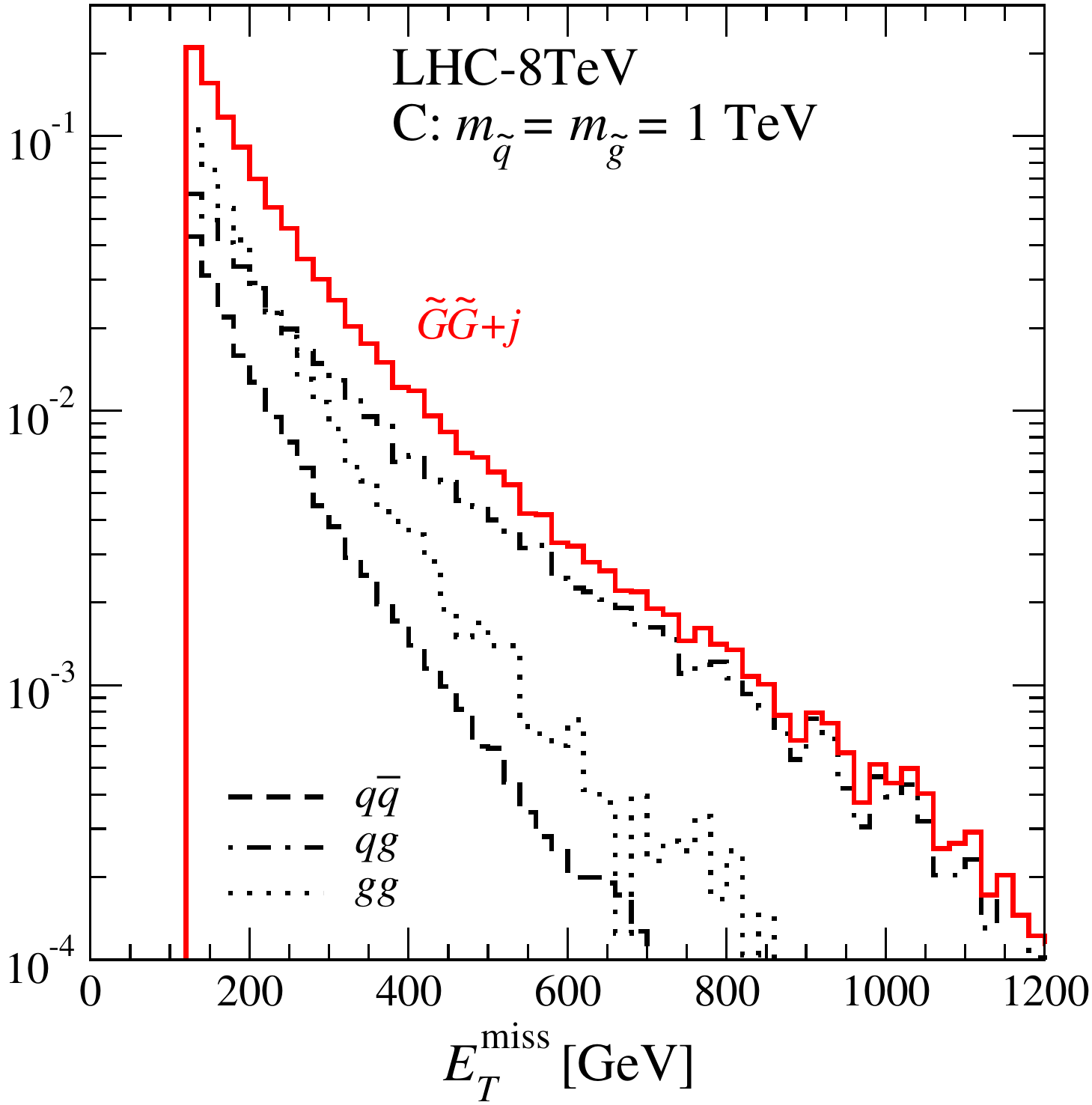}    
 \caption{Normalized missing transverse energy distributions of the
 direct gravitino-pair production with an extra radiation for the three
 benchmarks in~\eqref{cases} at the LHC-8TeV.
 Parton-shower and detector effects are included for the event
 generation and a cut $\Etmiss>120$~GeV is imposed.
 The contributions from different initial states are also shown.
 }
\label{fig:met_gldpair}
\end{figure}

\begin{figure}
\center
 \includegraphics[width=.495\textwidth,clip]{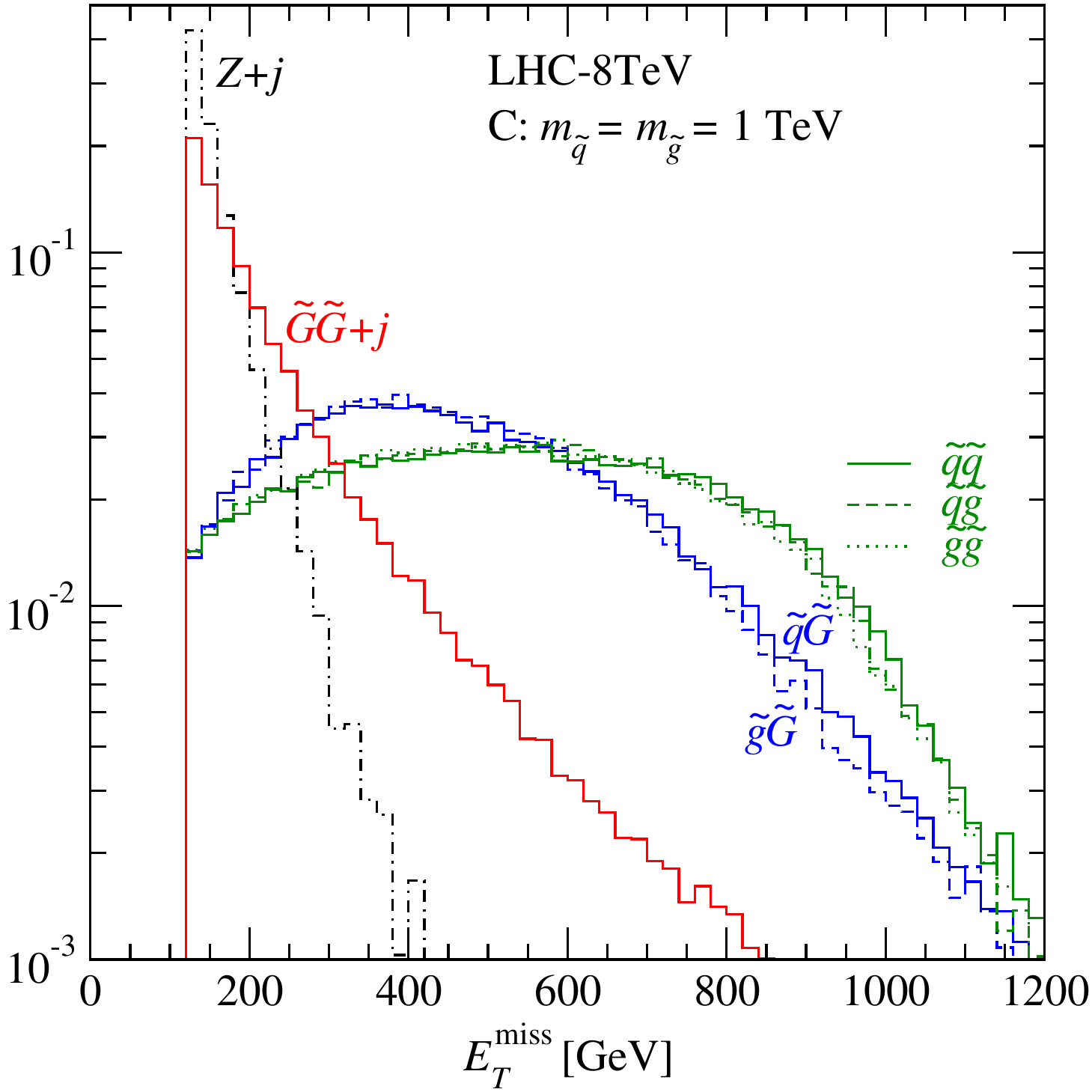} 
 \includegraphics[width=.495\textwidth,clip]{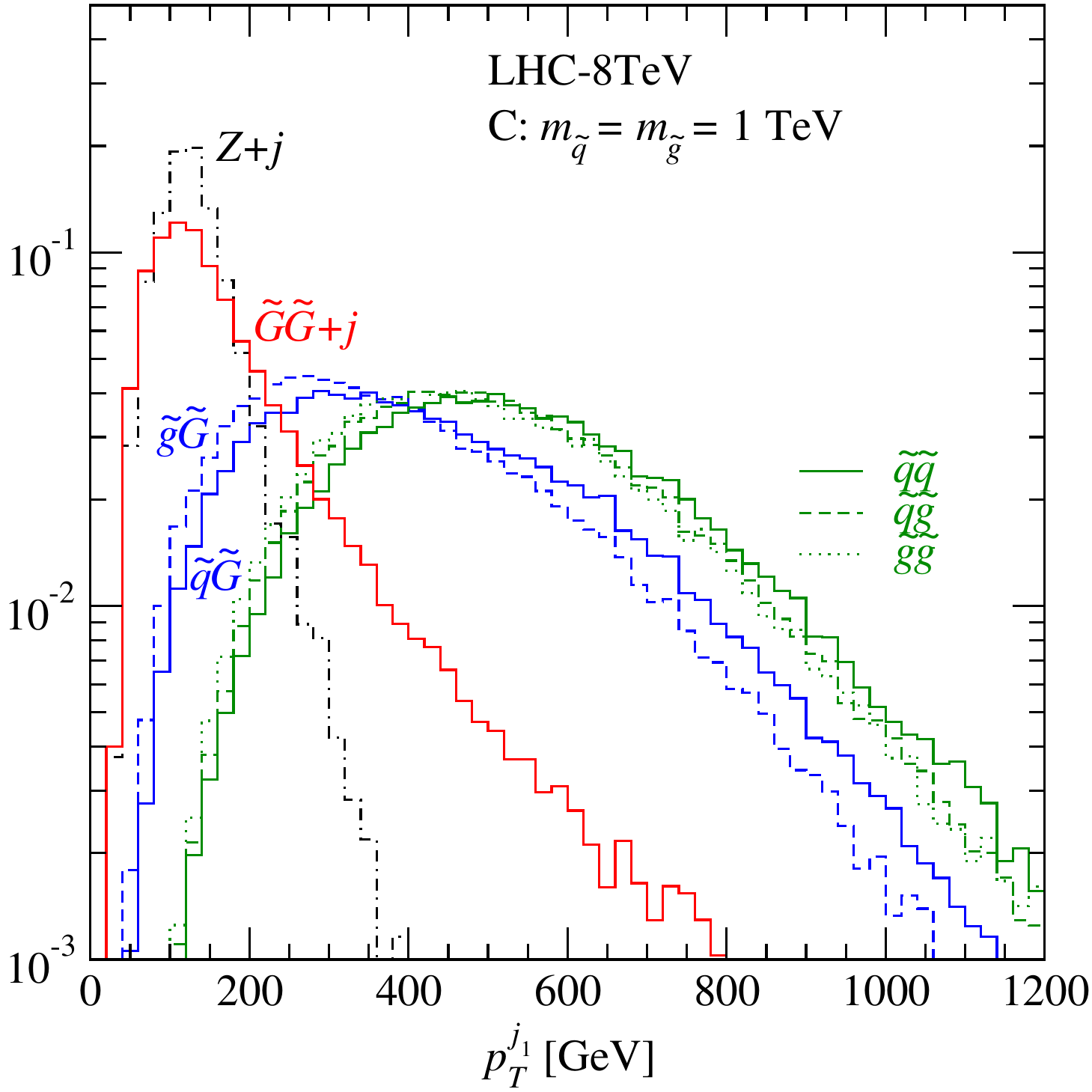}\\[2mm] 
 \includegraphics[width=.495\textwidth,clip]{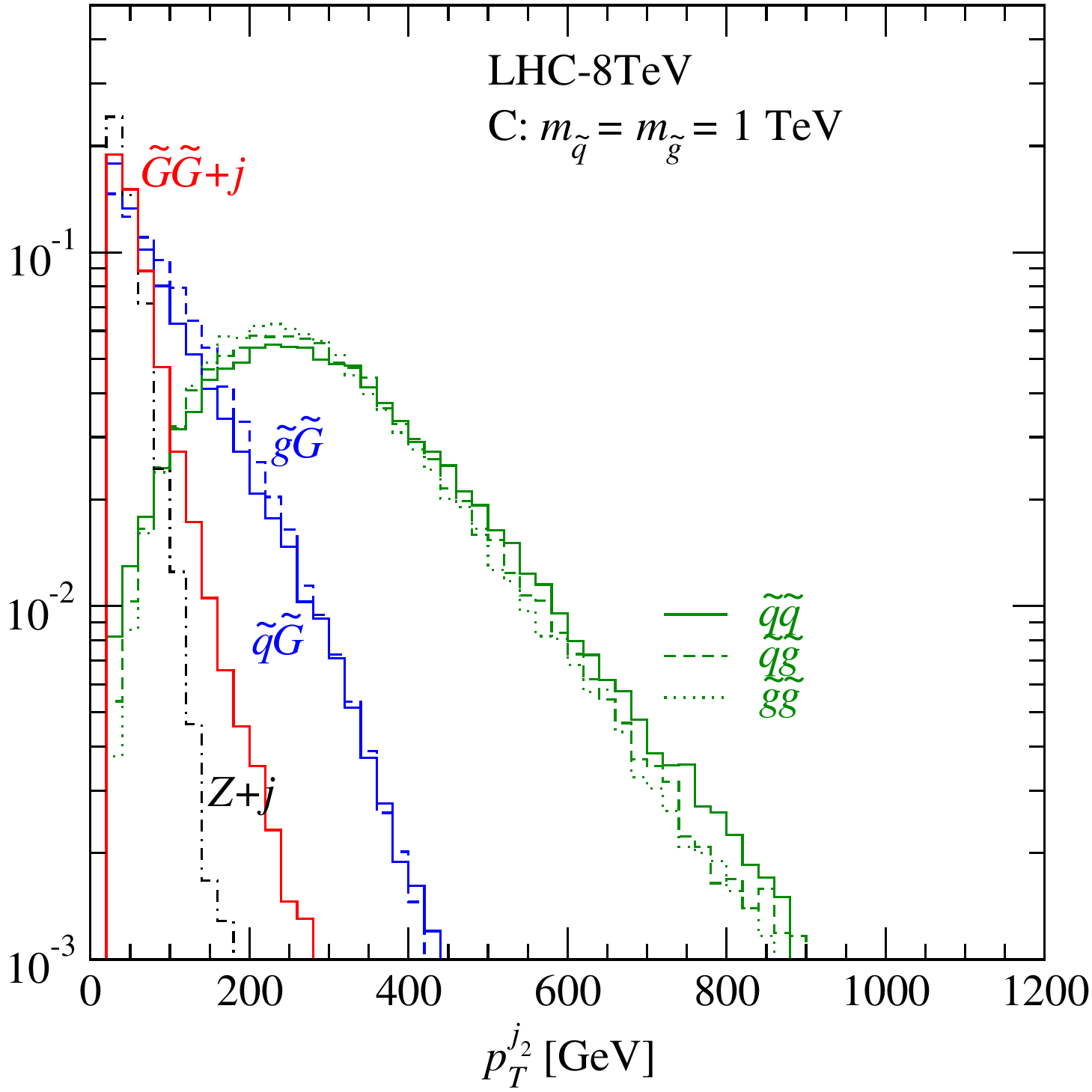}\   
 \includegraphics[width=.48\textwidth,clip]{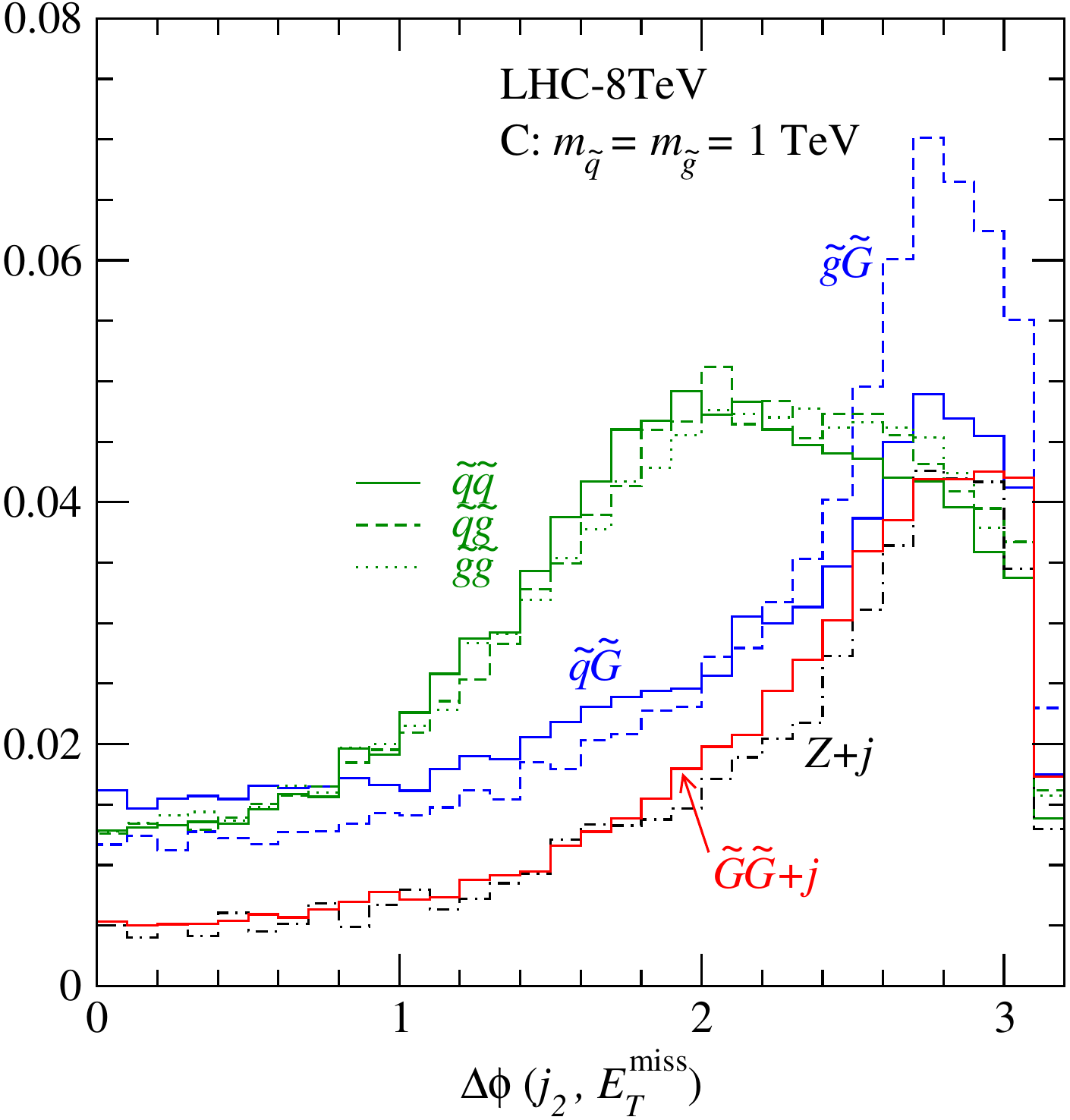} 
 \caption{Normalized distributions for each signal
 subprocess with 
 $m_{3/2}=2\times10^{-13}$~GeV and $m_{\sq,\go}=1$~TeV at the LHC-8TeV.
 Parton-shower and detector effects are included for the event
 generation, and a cut $\Etmiss>120$~GeV is imposed.
 As a reference, the $Z(\to\nu\bar\nu)+j$ background is also shown.}
\label{fig:dis}
\end{figure}

We now consider differential distributions for the direct
gravitino-pair production in detail. This is the first presented result that goes 
beyond the gravitino EFT limit. 
Figure~\ref{fig:met_gldpair} shows normalized missing transverse momentum
distributions for the three benchmark scenarios in~\eqref{cases}.
Here parton-shower and detector effects are included, and the detector
acceptance cuts $p_T^j>20$~GeV and $|\eta^j|<4.5$ as well as the missing
transverse momentum cut $\Etmiss>120$~GeV are applied.
Jets are reconstructed employing the anti-$k_T$
algorithm~\cite{Cacciari:2008gp} with a radius parameter of 0.4.
Depending on the mass of the $t$-channel exchanged squarks and gluinos,
the contributions from different initial states can be of different relevance.
Moreover, the energy spectra from $q\bar q$ and $gg$ are similar,
while that from $qg$ is harder than the others.

Figure~\ref{fig:dis} presents several kinematical distributions of
all the production channels for case
C as well as the SM $Z+j$ background. We stress that the purpose of including
the $Z+j$ background is illustrative on the one hand and to provide a 
``normalisation''
point for experimentalists. Needless to say, many other important sources of backgrounds
need to be included for a complete analysis, such as those coming from $W$+jets or just (mis-measured) jets. 
Most of them, however, can only be meaningfully estimated in presence of a 
detailed detector simulation and data validation. 

We see that the SUSY signals are harder than the
SM background, even for the gravitino-pair production. This is mainly
due to the $2\to3$ kinematics of the 
signal, whereas the background essentially has the $2\to2$ kinematics.  
Besides the background,
$\gld\gld(+j)$ has the softest spectra,
while $\sq\sq/\sq\go/\go\go$ lead the hardest.
The differences in the $p_T$ spectrum of the second-leading jet are rather significant.
The second jet mostly comes from the squark or gluino decay for
the SUSY QCD pair production, while mainly from QCD radiation
in the gravitino-pair and associated productions.   
We note that the shapes for the available subprocesses are very similar
among the three scenarios in~\eqref{cases}, while the rates are different as seen in the 
previous subsection.

\subsection{Recasting LHC mono-jet analyses}

ATLAS and CMS have reported a search for new physics in mono-jet plus
missing transverse momentum final states. 
The null results are translated into limits on a gauge-mediated SUSY,
large extra dimension and dark matter models in
ATLAS~\cite{ATLAS:2012zim} and 
on dark matter, large extra dimension and unparticle models in
CMS~\cite{Khachatryan:2014rra}.
As mentioned in the introduction, in the ATLAS analysis, a light
gravitino scenario has been studied, but only the squark-gravitino and gluino-gravitino
associated productions. 
In this section, taking into account all the possible gravitino
production processes described above, we recast the ATLAS 8-TeV mono-jet
analysis with 10.5~fb$^{-1}$ data~\cite{ATLAS:2012zim} to constrain the
gravitino mass for different squark and
gluino masses.

\subsubsection{Selection cuts}

The event selection of the ATLAS analysis~\cite{ATLAS:2012zim} is
\begin{align}
 &1.\quad \Etmiss>120~{\rm GeV}, \nn\\
 &2.\quad \text{leading jet with $p_T^{j_1}>120$~GeV and $|\eta^{j_1}|<2.0$,} \nn\\
 &3.\quad \text{at most two jets with $p_T^{j}>30$~GeV and
 $|\eta^{j}|<4.5$,} \nn\\
 &4.\quad \Delta\phi(j_2,\Etmiss)>0.5.
\label{cuts}
\end{align}
The third requirement allows the second-leading jet ($j_2$) since signal
events  typically contain jets from initial state radiation, while the last one
reduces the QCD background where the large $\Etmiss$ originates from the
mis-measurement of $p_T^{j_2}$.
On top of the above requirements, similarly to the ATLAS analysis, we define three signal regions (SRs) with
different $\Etmiss$ and $p_T^{j_1}$ thresholds as%
\footnote{SR2 in the ATLAS analysis is with the 220~GeV
cut~\cite{ATLAS:2012zim}. On the other hand, 
our SR2 is similar to the one of the signal regions in the CMS 
analysis~\cite{Khachatryan:2014rra}.}
\begin{align}
 &{\rm SR1}:\quad \Etmiss, p_T^{j_1}>120~{\rm GeV}, \nn\\
 &{\rm SR2}:\quad \Etmiss, p_T^{j_1}>250~{\rm GeV}, \nn\\
 &{\rm SR3}:\quad \Etmiss, p_T^{j_1}>350~{\rm GeV}.
\label{sr}
\end{align}

\begin{table}
\center
\begin{scriptsize}
\begin{tabular}{ll||r|rrr|rrrrrr|r}
\hline
 && \multicolumn{1}{c|}{A} & \multicolumn{3}{c|}{B} 
  & \multicolumn{6}{c|}{C} & \multicolumn{1}{c}{bkg} \\
 && $\gld\gld$ 
  & $\gld\gld$ & $\go\gld$ & $\go\go$ 
  & $\gld\gld$ & $\sq\gld$ & $\go\gld$ & $\sq\sq$ & $\sq\go$ & $\go\go$
  & $Z+j$\\
\hline\hline 
 & $\quad\Etmiss>120$~GeV
  & 5257
  & 5433 & 1770 & 140
  & 1400 & 878 & 353 & 1716 & 938 & 79 
  & 329893\\ 
 & $+\,p_{T}^{j_1}>120$~GeV 
  & 3164 
  & 3291 & 1672  & 139
  & 800 & 836 & 336 & 1698 & 929 & 79 
  & 163270\\
 & $+\,$at most 2 jets
  & 2776 
  & 2869 &1108 & 15 
  & 614 & 550 & 180 & 589 & 138 & 6
  & 152532\\
 SR1 & $+\,\Delta\phi(j_2,\Etmiss)>0.5$ 
  & 2690
  & 2778 & 1061 & 14 
  & 583 & 508 & 170 & 551 & 128 & 5 
  & 146548\\
 SR1' & $+\,p_T^{j_2}<150$~GeV
  & 2652
  & 2736 & 959 & 3 
  & 564 & 455 & 152 & 88 & 23 & 1
  & 145954\\
\hline\hline
 & SR1$\,+\Etmiss>250$~GeV 
  & 869 
  & 914 & 956 & 13 
  & 229 & 454 & 153 & 497 & 116 & 5 
  & 12604\\
 SR2 & $+\,p_{T}^{j_1}>250$~GeV 
  & 614 
  & 654 & 863 & 12 
  & 170 & 424 & 138 & 487 & 114 & 5
  & 7554\\
 SR2' & $+\,p_{T}^{j_2}<150$~GeV
  & 591
  & 628 & 778 & 2 
  & 157 & 379 & 123 & 75 & 21 & 1 
  & 7512\\
\hline\hline
 & SR2$\,+\Etmiss>350$~GeV
  & 340
  & 369 & 762 & 11 
  & 109 & 361 & 120 & 432 & 102 & 4 
  & 2037\\
 SR3 & $+\,p_{T}^{j_1}>350$~GeV
  & 254  
  & 281 & 660 & 10 
  & 86 & 323 & 103 & 403 & 94 & 4 
  & 1358\\
 SR3' & $+\,p_{T}^{j_2}<150$~GeV
  & 243 
  & 268 & 604 & 2 
  & 79 & 291 & 93 & 61 & 17 & 1 
  & 1358\\
\hline
\end{tabular}
\end{scriptsize}
\caption{SUSY signal predictions of the three scenarios in~\eqref{cases}
 with $m_{3/2}=2\times10^{-13}$~GeV for the number of events passing 
 each step of the
 selection requirements in~\eqref{cuts} and \eqref{sr}, expected for an integrated luminosity of 10.5~fb$^{-1}$
at
 the LHC-8TeV.
 $Z(\to\nu\bar\nu)+j$ background is also shown as a reference.}
\label{tab:cutflow}
\end{table}

In table~\ref{tab:cutflow} we present SUSY signal predictions for the
number of events passing each step of the above selection requirements.
As in fig.~\ref{fig:dis}, we generate events for each subprocess
including parton-shower and detector effects. 
In addition to the three SUSY benchmark scenarios in~\eqref{cases} with
the gravitino mass at $2\times10^{-13}$~GeV, we show the 
$Z(\to\nu\bar\nu)+j$ background prediction, which is the dominant
background, as a reference; see table~2 in the ATLAS analysis~\cite{ATLAS:2012zim} for more
details on the background estimation including other channels.

At the LO parton level, $\Etmiss=p_T^{j_1}$ for the $\gld\gld(+j)$
and $\sq\gld/\go\gld$ productions. After the parton shower, the
relation does not hold any more, and the effect of the radiation is quite large for the gravitino
pair production. 
As expected, the third selection cut in~\eqref{cuts} does not affect so
much for $\gld\gld(+j)$ and $\sq\gld/\go\gld$, while
significantly reduces the SUSY QCD pair contributions 
although for case C the contribution is still substantial and even
dominant in SR3. 
We remind the reader that  SUSY QCD pair production is insensitive
to the gravitino mass if the gravitino is heavier than
$3\times10^{-13}$~GeV,
and hence these contributions have to be considered as background to constrain the
gravitino mass.
To reduce this SUSY QCD background, on top of the above signal selection
cuts, we impose a maximal $p_T$ cut
on the second-leading jet in each SR as
\begin{align}
 p_T^{j_2}<150~{\rm GeV},
\label{ptj2cut}
\end{align}
denoted as SR1', SR2' and SR3'.
As can be also seen in the $p_T^{j_2}$ distribution in fig.~\ref{fig:dis}, 
this cut removes the large part of the events coming from $\sq\sq$,
$\sq\go$ and $\go\go$.

\subsubsection{Merging matrix elements with parton showers}

\begin{figure}
\center
 \includegraphics[width=.328\textwidth,clip]{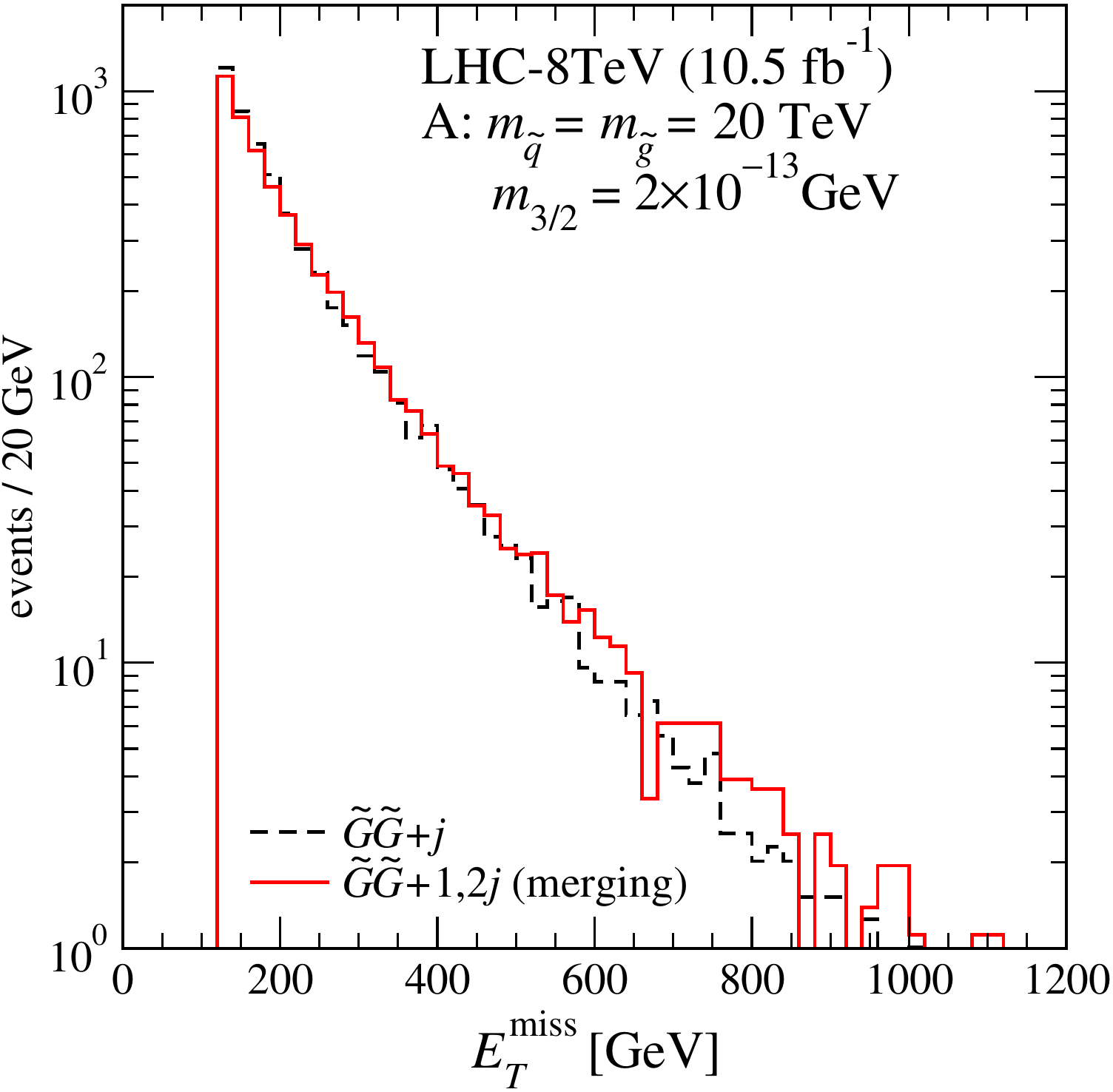}
 \includegraphics[width=.328\textwidth,clip]{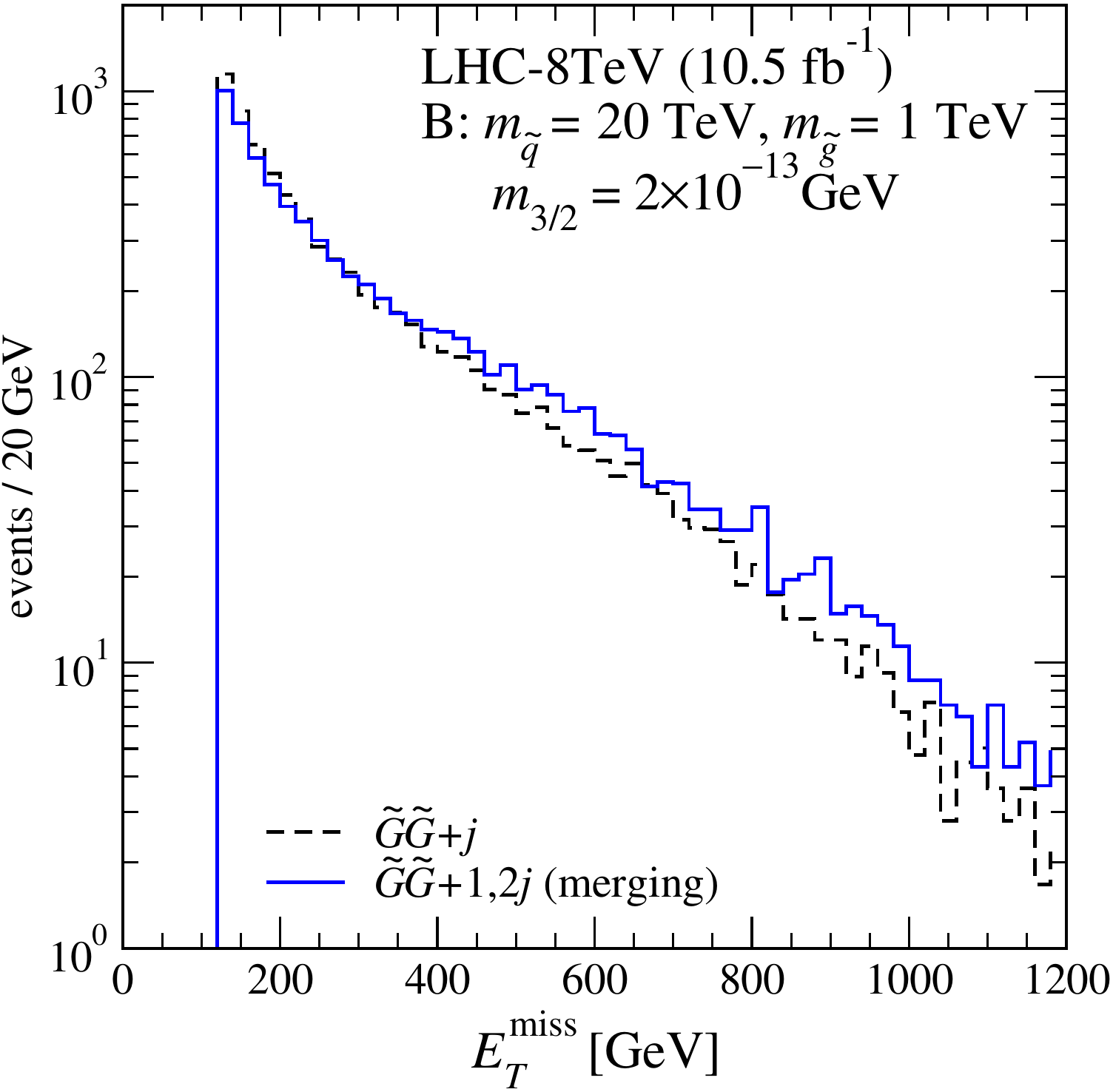}
 \includegraphics[width=.328\textwidth,clip]{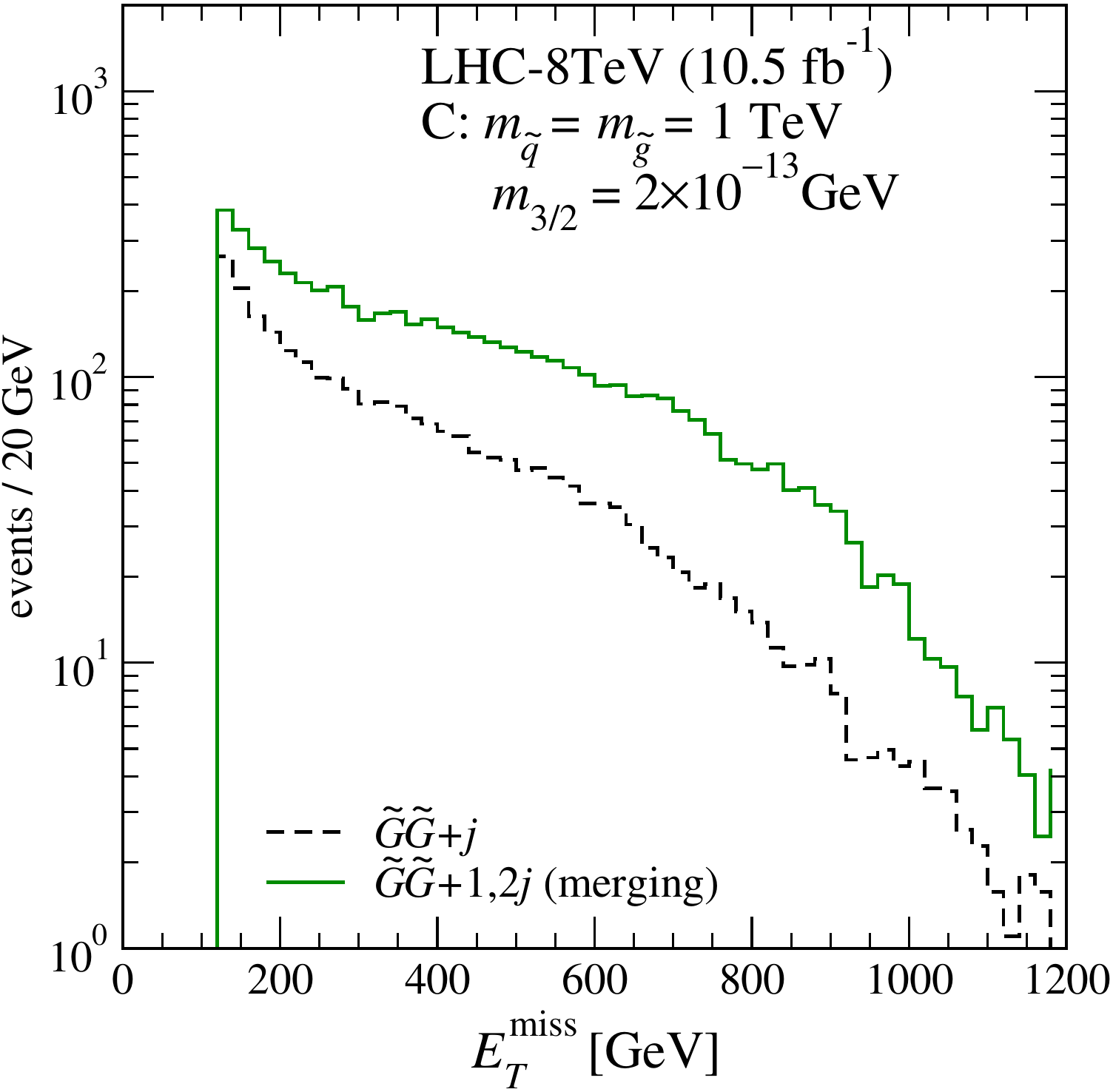}    
 \caption{Missing transverse energy distributions for the three
 scenarios in~\eqref{cases} with
 $m_{3/2}=2\times 10^{-13}$~GeV, where the inclusive samples of the
 $\gld\gld+1$ parton in the matrix element (dashed) are compared with the merged
 samples containing an extra parton (solid). 
 Only the cut $\Etmiss>120$~GeV is applied.
 }
\label{fig:met}
\end{figure}

So far, in order to identify characteristics and differences among them we have treated each gravitino-production subprocess independently.  Now, to constrain the SUSY mass parameters,
we generate inclusive signal samples 
by using the ME+PS merging procedure.
In practice, following ref.~\cite{deAquino:2012ru}, we make use of the
shower-$k_T$ scheme~\cite{Alwall:2008qv}, and  
generate signal events with parton multiplicity from one
to two, $pp\to\gld\gld+1,2$ partons, and merging separation parameters
$Q_{\rm cut}=60$~GeV and $p_{T_{\rm min}}=50$~GeV.
We checked carefully that the variation of $Q_{\rm cut}$ did not
change the distributions after the minimal missing transverse energy cut $\Etmiss>120$~GeV.
The factorization and renormalization scales are set to the scalar
sum of the $p_T$ of all the partons in the final state.
We note that the employment of
the ME+PS merging procedure allows us to treat different contributing
processes, i.e. gravitino-pair, associated gravitino and SUSY QCD pair
productions (see also fig.~\ref{fig:diagram}),  
within one event simulation and without double counting.
We also note that the interference among the different production
processes is very small since the width of the on-shell squarks and
gluinos is small with our parameter choice.

To see the effect of an extra parton in the matrix element, 
in fig.~\ref{fig:met} we compare the inclusive samples of the
$\gld\gld+1$ parton in the matrix element with the merged samples of 
$pp\to\gld\gld+1,2$ partons.
For case A, where only the gravitino-pair production contributes, we find
a slightly harder spectrum in the high $\Etmiss$ region for the
merged sample due to the second parton in the matrix element. 
For case B, as seen in table~\ref{tab:cutflow}, besides
$\gld\gld$, the $\go\gld$ production contributes significantly,
leading to a much harder spectrum than in case A.  
Again, a harder spectrum for the merged sample is observed as expected.
For case C, with the minimal selection cut $\Etmiss>120$~GeV, the SUSY
QCD pair productions, especially $\sq\sq$ and $\sq\go$, are dominant,
which do not exist in the $\gld\gld+1$ parton sample.
Therefore, the distributions are completely different without and with
an extra parton in the matrix-element level.

\subsubsection{Limit on the gravitino mass}
 
\begin{figure}
\center
 \includegraphics[width=.495\textwidth,clip]{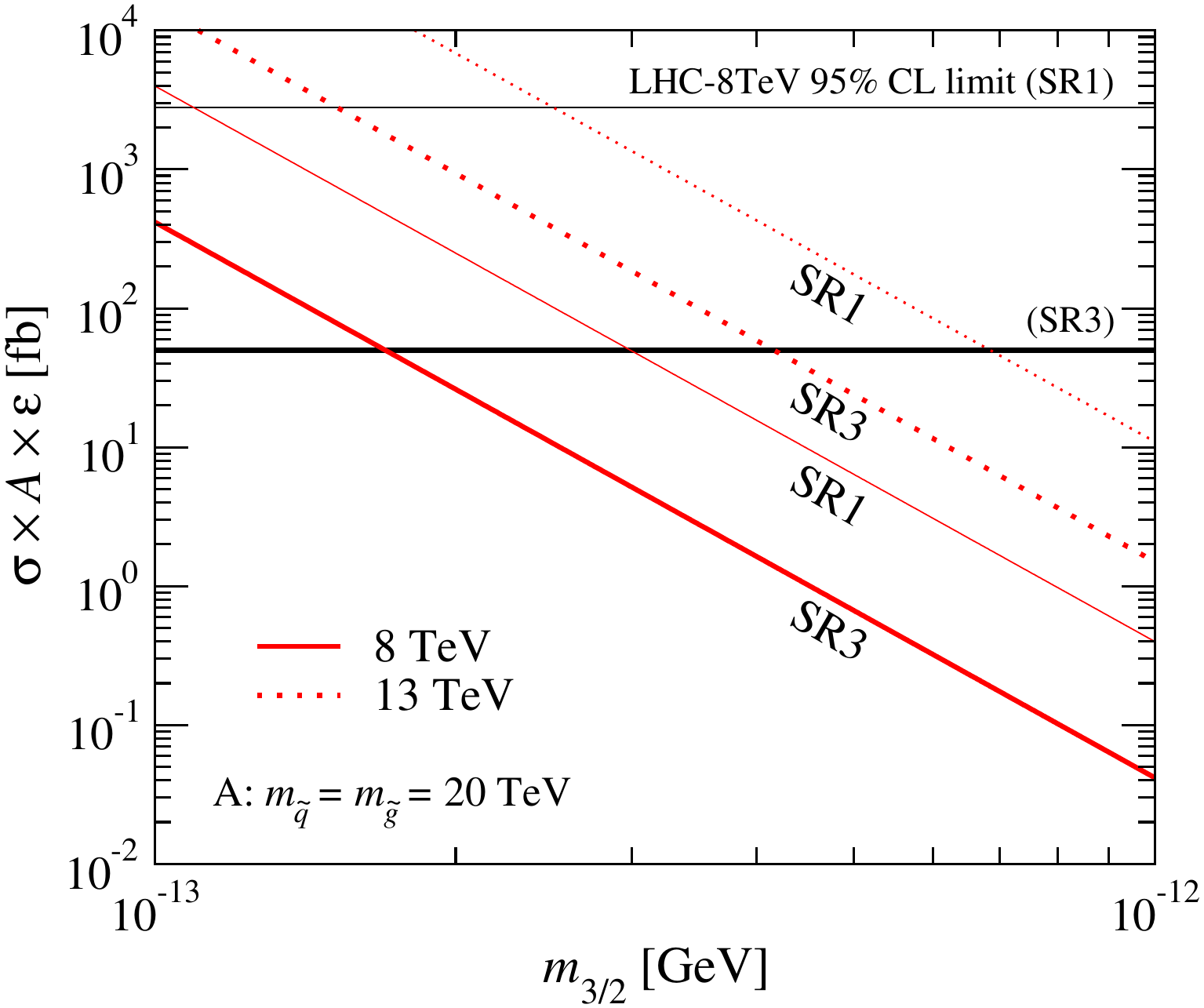} 
 \includegraphics[width=.495\textwidth,clip]{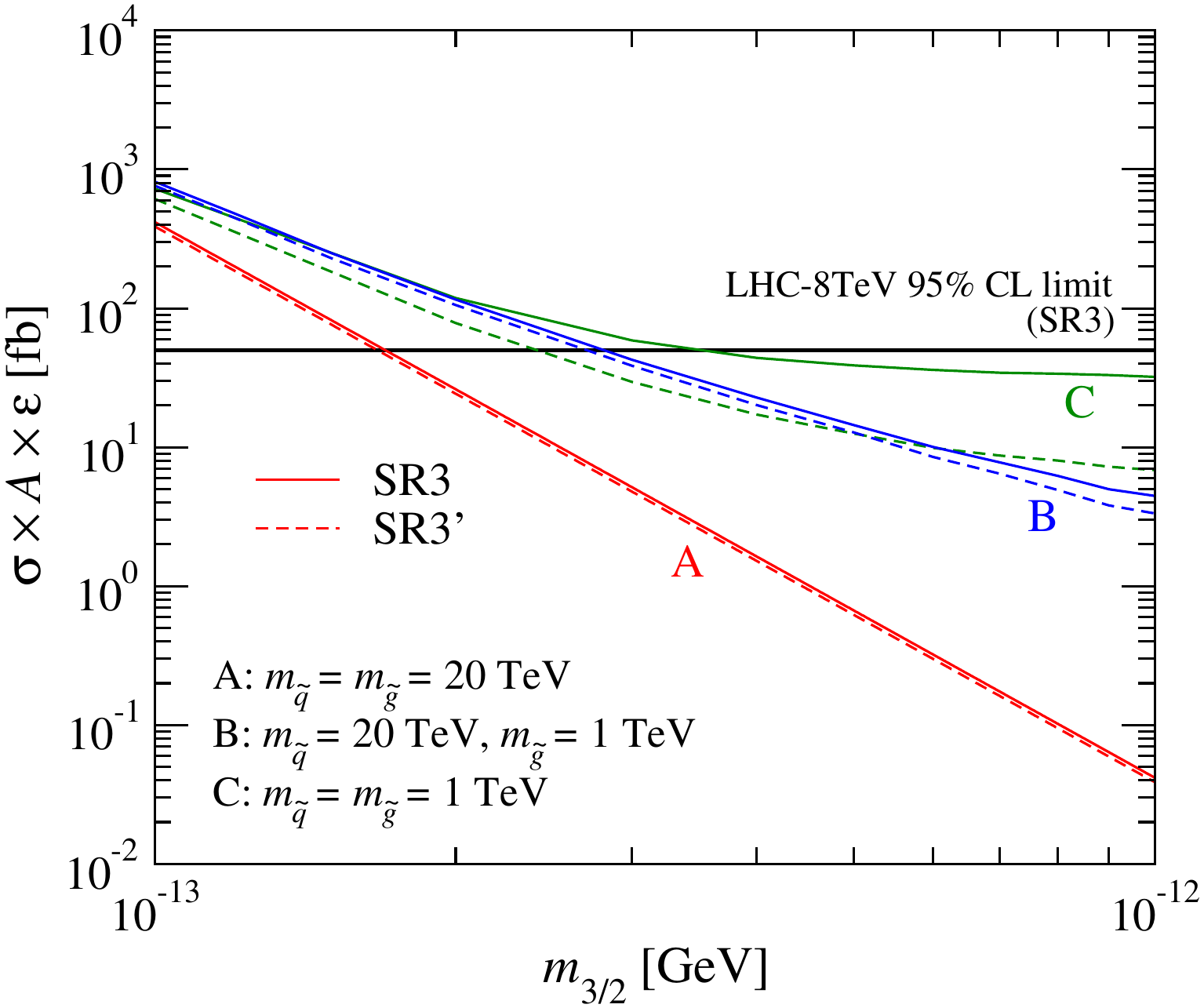} 
 \caption{Left: Visible cross sections of the mono-jet
 signal for case A at $\sqrt{s}=8$~TeV (solid) and 13~TeV (dotted) as a
 function of the gravitino mass, where SR1 and SR3 are shown. 
 The predictions are compared with the model-independent 95\% confidence level (CL) upper
 limits by the ATLAS analysis~\cite{ATLAS:2012zim}.
 Right: Same as the left panel, but for all the three scenarios in SR3 (solid)
 and SR3' (dashed) at $\sqrt{s}=8$~TeV. 
 }
\label{fig:limit_j}
\end{figure}

By using the inclusive ME+PS merged samples, we can now recast the
ATLAS-8TeV mono-jet analysis with 10.5~fb$^{-1}$ data
set~\cite{ATLAS:2012zim}. ATLAS  
reported a model-independent 95\% confidence level (CL) upper limit on the visible cross
section, defined as the production cross section times kinematical
acceptance times detection efficiency ($\sigma\times A\times\varepsilon$). 
The values are $2.8\times10^3$~fb and 50~fb for SR1 and SR3 selections, respectively. 

Figure~\ref{fig:limit_j} (left) presents the visible cross sections for
case A at $\sqrt{s}=8$ and 13~TeV as a function of the gravitino mass.
The horizontal lines show the ATLAS 95\% CL limits.
In SR1 the SM background is huge, and hence only the very light
gravitino case can be constrained.
The constraint in SR3 is slightly better than in SR1, and the gravitino
mass below about $1.7\times10^{-13}$~GeV are excluded at 95\% CL in the
gravitino EFT. 
This limit is one order of magnitude stronger than the limits at the LEP and
Tevatron~\cite{Agashe:2014kda}.  
According to the relation in~\eqref{grav_mass}, the above limit
corresponds to the SUSY breaking scale of about 850~GeV.
The coming LHC Run-II with $\sqrt{s}=13$~TeV is expected to explore
heavier gravitinos up to ${\cal O}(10^{-12})$~GeV, i.e. a few TeV of the
SUSY breaking scale.

In fig.~\ref{fig:limit_j} (right), the visible cross sections in SR3 at
$\sqrt{s}=8$~TeV are
shown for case A, B and C. 
Roughly speaking, case A and B follow $m_{3/2}^{-4}$ and
$m_{3/2}^{-2}$, respectively, as expected. 
For case C, on the other hand, no sensitivity of the cross section to the
gravitino mass is observed when the gravitino mass is heavier than about
$3\times10^{-13}$~GeV.
However, by imposing an additional cut on the second-leading jet
in~\eqref{ptj2cut}, the sensitivity to the gravitino mass recovers even
for heavier gravitinos since the SUSY QCD pair productions are strongly
suppressed. 
The maximal $p_T^{j_2}$ cut hardly affects the signals for case A and B.

\section{Mono-photon, -$Z$, or -$W$ plus missing momentum}
\label{sec:photon}

\begin{figure}
\center
 \includegraphics[width=.7\textwidth,clip]{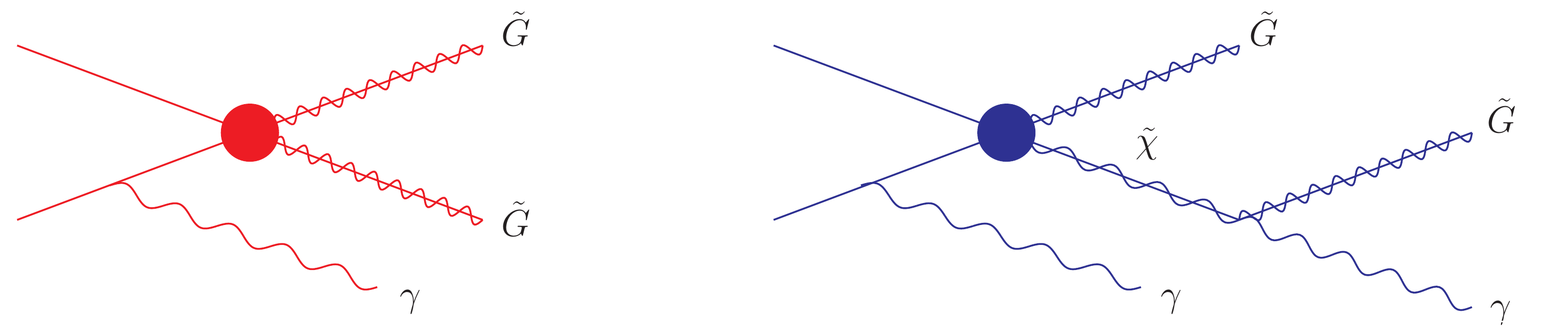} 
 \caption{Schematic diagrams for $pp\to\gld\gld+\gamma$, where 
 gravitino-pair production with a photon emission (left) and
 neutralino--gravitino associated production (right) contribute. 
}
\label{fig:diagram_a}
\end{figure}

In an analogous way to the mono-jet signal discussed in the previous section, 
superlight gravitino scenarios can provide mono-$\gamma$, -$Z$, or -$W$ 
(mono-EW boson) plus missing momentum signature via
\begin{enumerate}
\item gravitino-pair production with a $\gamma$, $Z$, or $W$ emission,
\item gravitino production associated with a neutralino/chargino with the
      subsequent decay into a $\gamma,Z/W$ and a gravitino.
\end{enumerate}
The schematic diagrams are shown in fig.~\ref{fig:diagram_a}.
Unlike the $j+\Etmiss$ signal, only the $q\bar q$ initial state can
contribute to the mono-EW boson$+\Etmiss$ signal.
In this section, for simplicity, we consider the heavy
neutralino/chargino limit, where only the gravitino-pair production
contributes.%
\footnote{The mono-photon signal of $\tilde\chi^0_1\gld$ production via
the Higgs decay at the LHC was studied in~\cite{Petersson:2012dp}.}   

\begin{figure}
\center
 \includegraphics[width=.495\textwidth,clip]{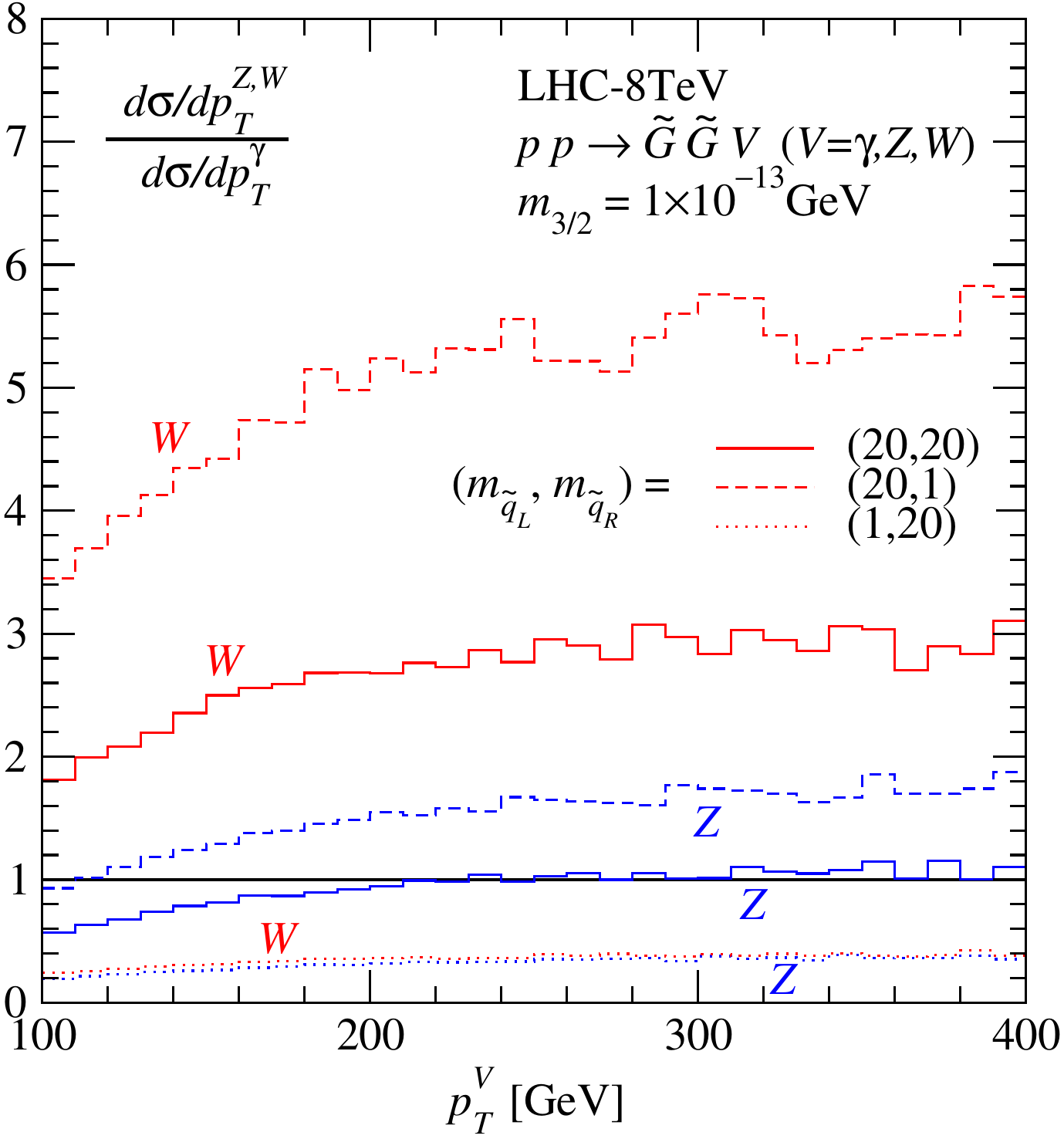} 
 \caption{Ratio of the transverse momentum distributions of the $Z$ or $W$
 boson to that of the photon for 
 $pp\to\gld\gld V$ $(V=\gamma,Z,W)$ at $\sqrt{s}=8$~TeV with
 $m_{3/2}=1\times10^{-13}$~GeV.
 Three scenarios for different left- and right-handed squark masses are
 considered.} 
\label{fig:pt_azw}
\end{figure}

So far, new physics searches in mono-$\gamma$, -$Z$ and -$W$ signals
at the LHC have been done independently, but the combined analysis may be very
interesting because there is a possibility to determine left--right
handedness of the new physics interactions.
Instead of studying the gravitino-mass constraint in each search
channel,  fig.~\ref{fig:pt_azw} shows the ratio of the $p_T$
distributions of the massive gauge boson to that of the photon for 
$pp\to\gld\gld V$ $(V=\gamma,Z,W)$ at $\sqrt{s}=8$~TeV with
$m_{3/2}=1\times10^{-13}$~GeV.
There are $t$-channel squark exchange diagrams, and for illustration we take three left-
and right-handed squark mass scenarios: 
\begin{align}
 (m_{\sq_L},m_{\sq_R})=\{(20,20),\, (20,1),\, (1, 20)\}~{\rm TeV}. 
\end{align}
The effect of the mass of the gauge boson can be seen as suppression
and enhancement in the low and high $p_T$ region, respectively. 
Interestingly, the ratios are very sensitive to the mass difference between
$\sq_L$ and $\sq_R$, especially for the $W$ boson, which only couples to
the left-handed squarks.  

Finally, we recast the LHC-8TeV mono-photon
analyses~\cite{Khachatryan:2014rwa,Aad:2014tda}, where non-SUSY models
were studied, to constrain the gravitino mass. 
For event selection, we follow the $\gamma+\Etmiss$ analysis by
ATLAS~\cite{Aad:2014tda}. 
Events in the signal region are required to have 
the missing transverse energy
$\Etmiss>150$~GeV and  
a photon with
$p_T>125$~GeV and $|\eta|<1.37$. 
The photon and the missing momentum
vector are also required to be well separated as
$\Delta\phi(\gamma,\Etmiss)>0.4$.
Possible jets produced by ISR are
defined by the anti-$k_T$ algorithm~\cite{Cacciari:2011ma} with a 
radius parameter of 0.4 and are required to be in the region 
$|\eta|<4.5$ with $p_T>30$~GeV. 
While events with more than one
jet are rejected,
events with one jet with $\Delta R(\gamma,j)>0.2$ and
$\Delta\phi(\Etmiss,j)>0.4$ are kept for the signal with ISR, where
$\Delta R=\sqrt{(\Delta\eta)^2+(\Delta\phi)^2}$.

\begin{figure}
\center
 \includegraphics[width=.5\textwidth,clip]{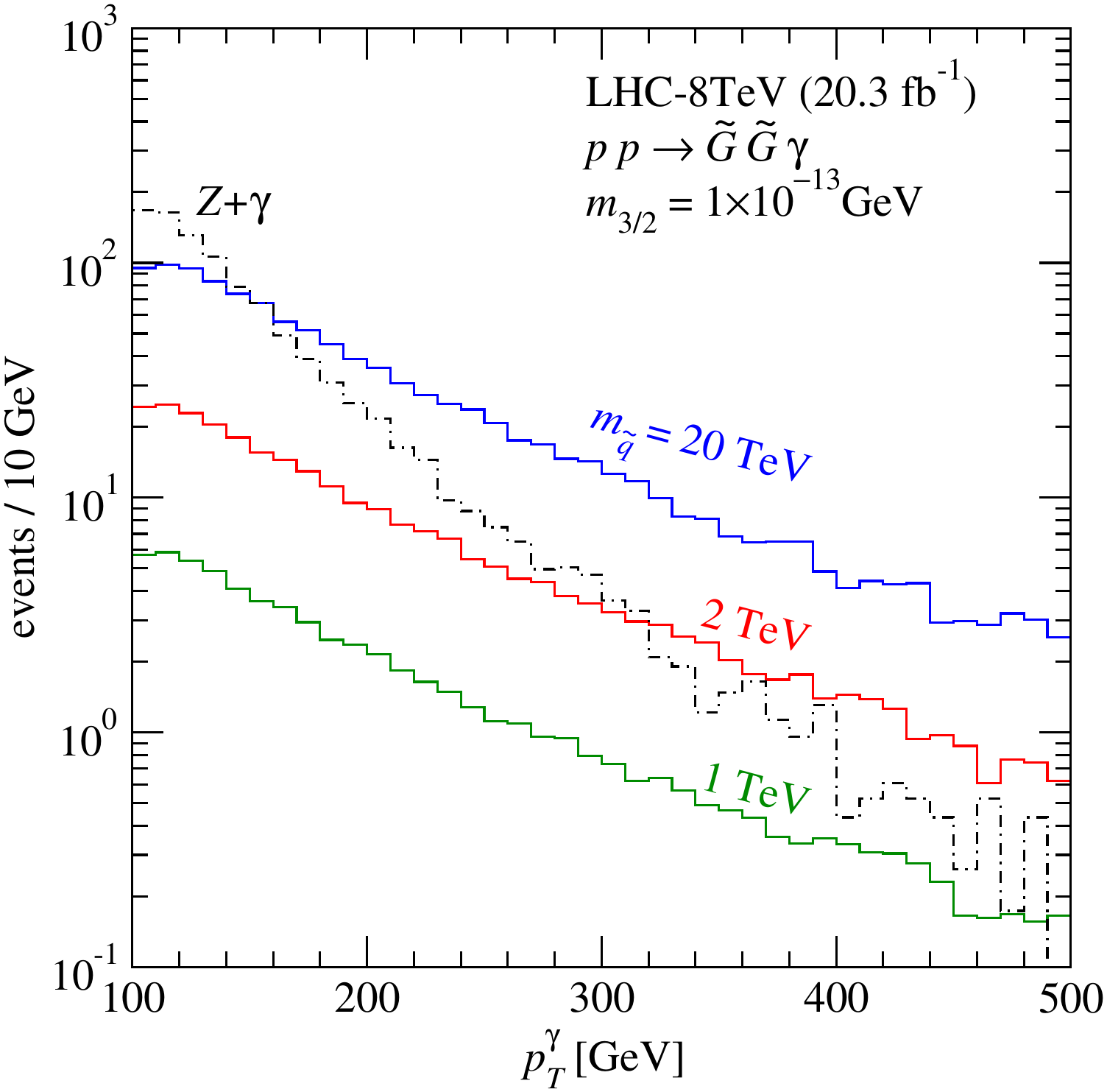} 
 \caption{Transverse momentum distributions of the photon for 
 $pp\to\gld\gld\gamma$ at $\sqrt{s}=8$~TeV with $m_{3/2}=1\times10^{-13}$~GeV
 for three squark masses. All selection cuts described in the text are
 applied except the $p_T^{\gamma}$ and $\Etmiss$ cuts. The 
 $Z(\to\nu\bar\nu)+\gamma$ background is also shown as a reference.}
\label{fig:pt_a}
\end{figure}

Figure~\ref{fig:pt_a} shows the $p_T$ distributions of the photon for
$pp\to\gld\gld\gamma$ at $\sqrt{s}=8$~TeV, where all the above selection
cuts are applied except the $p_T^{\gamma}$ and $\Etmiss$ cuts. The
gravitino mass is fixed at $1\times10^{-13}$~GeV, while the masses of squarks
are taken at 1, 2, and 20~TeV. As discussed in the mono-jet signal, the
cross section for the gravitino-pair production becomes larger as the
$t$-channel squark masses increase. 
In analogy with the mono-jet case, the SUSY signal is harder than the
SM background mainly due to the kinematics.
We note again that the signal rate strongly
depends on the gravitino mass as $m_{3/2}^{-4}$ and also on the
kinematical cuts.

\begin{figure}
\center
 \includegraphics[width=.5\textwidth,clip]{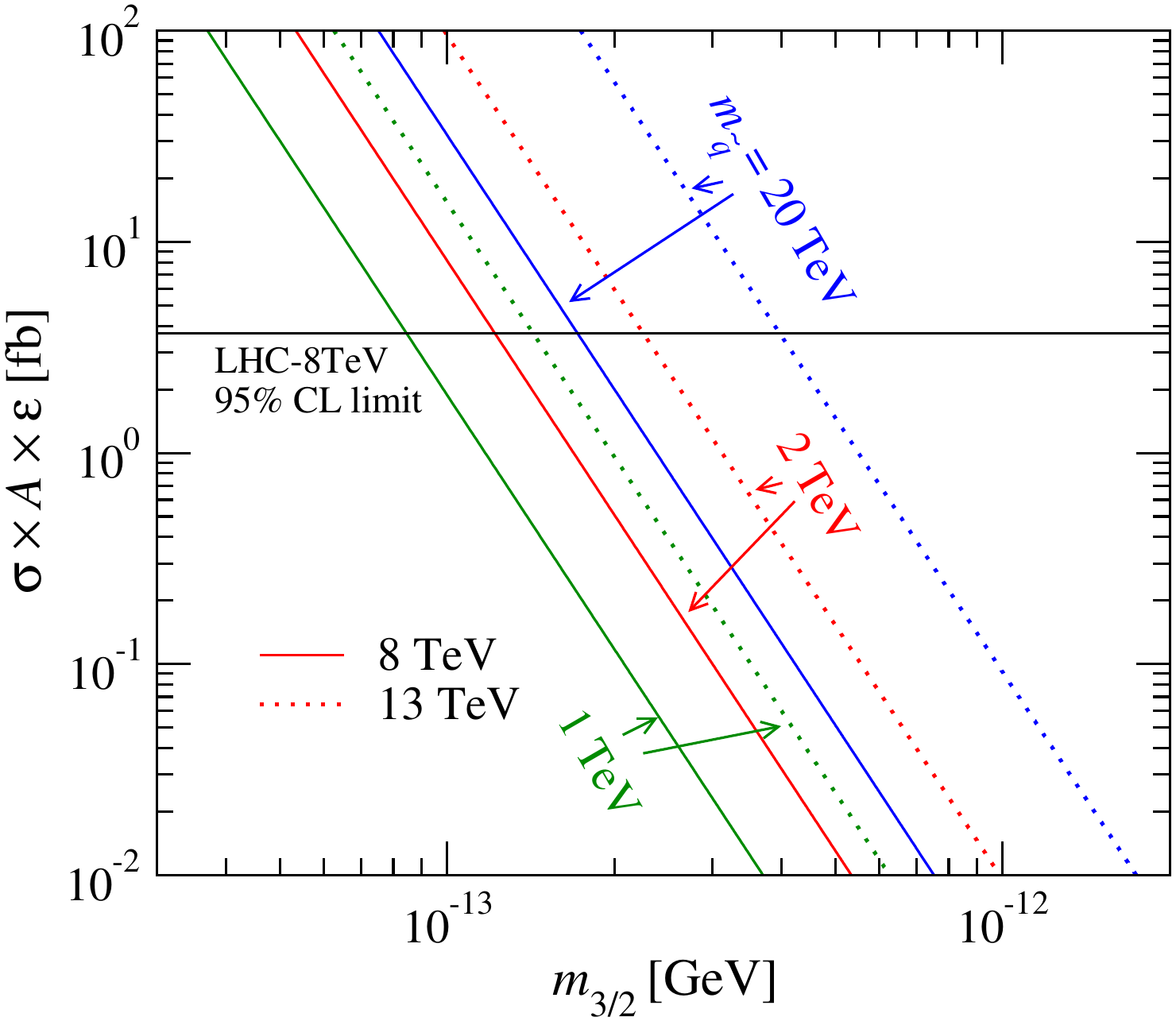} 
 \caption{Visible cross sections of the mono-photon signal at
 $\sqrt{s}=8$~TeV (solid) and 13~TeV (dotted) as a function of the
 gravitino mass for different squark masses.
 The predictions are compared with the
 model-independent 95\% confidence level (CL) upper limit by
 the ATLAS analysis~\cite{Aad:2014tda}.}
\label{fig:xsec_a}
\end{figure}

The ATLAS $\gamma+\Etmiss$ study with 20.3~fb$^{-1}$ of collisions at
$\sqrt{s}=8$~TeV reported a model-independent 95\% CL
upper limit on the fiducial cross section,
$\sigma\times A$. The value is 5.3~fb~\cite{Aad:2014tda}. 
Figure~\ref{fig:xsec_a}
presents the visible cross sections for $pp\to\gamma\gld\gld$ at
$\sqrt{s}=8$ and 13~TeV as a function of the gravitino mass for three
different squark masses. The horizontal line shows the ATLAS 
95\% CL limit, where we take a conservative estimate for
the fiducial reconstruction efficiency $\varepsilon=0.7$~\cite{Aad:2014tda}. 

Gravitino masses below about $1.7\times 10^{-13}$~GeV are excluded at 95\%
CL for the heavy SUSY mass limit, which is translated to the lower bound
on the SUSY breaking scale of about 850~GeV, similar to the mono-jet
limit. For lighter squark masses the 
limits are lower, for example, $m_{3/2}\sim 8.4\times 10^{-14}$~GeV,
i.e. $\sqrt{F}\sim 600$~GeV for 1-TeV squarks. These results
significantly improve previous ones at LEP and the Tevatron, and are
comparable with the recent ATLAS 8-TeV mono-jet 
analysis~\cite{ATLAS:2012zim}.%
\footnote{In the ATLAS study, only associated gravitino production with
a gluino or a squark was considered.}
The coming LHC Run-II with $\sqrt{s}=13$~TeV is expected to explore
heavier gravitinos up to ${\cal O}(10^{-12})$~GeV, i.e. a few TeV of the
SUSY breaking scale. We note that we assumed the heavy neutralino limit
in this section. However, if the neutralino is light enough and promptly
decays, production of the on-shell neutralino can give rise to
characteristic harder photons. This leads to different production rate
as well as $A\times\varepsilon$, and hence the limits can be
modified. The discussions for the mono-jet study in the previous section
can be applied for the mono-photon case by the replacement of
gluino/gluon to neutralino/photon for the $q\bar q$ initial state. 

\section{Summary}\label{sec:summary}

The mono-jet plus missing momentum signal at the LHC is a promising final
state where to look for new physics. In this work we investigated
the possibility of observing a SUSY signal via a very
light gravitino. Gravitino-pair production with extra radiation and
associated gravitino production with a squark or a gluino contribute both to 
mono-jet signals.  Moreover, in the current ATLAS and
CMS mono-jet analyses, squark and gluino pair production may contribute
to the signal region. We have carefully investigated the impact of consistently including  all three
production channels. We have constructed a SUSY QCD model, lifting previous limitations of
gravitino-EFT models. We have implemented it in the {\sc{FeynRules}} and {\sc{MadGraph5\_aMC@NLO}}
simulation framework paying special attention to needed Majorana four-fermion interactions.  

We discussed the parameter dependence of the signal rate in detail and
showed  that the relative importance of the three contributing subprocesses
varies with the gravitino and SUSY particle masses. We also studied the differential distributions to get better understanding of the expected shape for different parameters.

To constrain the gravitino and other SUSY masses we have recast the
LHC-8TeV mono-jet analyses by the ATLAS and CMS collaborations. 
Using matrix-element/parton-shower merged samples, we have been able to treat
all three contributing subprocesses within one event simulation and
without double counting. Re-interpreting the reported model-independent 95\%
CL upper limit on the visible cross section, we found that
a gravitino mass below about $1.7\times10^{-13}$ GeV is excluded in the
limit where all SUSY particles except the gravitino are very
heavy. We showed that this limit changes when allowing squarks and gluinos to be relatively
light. To get a better sensitivity to the
gravitino mass, we suggest an additional cut in the analysis which suppresses
contributions from SUSY QCD pair production. We have also discussed  prospects
for the LHC Run-II, which is expected to explore gravitino masses up to $\mathcal{O}(10^{-12})$ GeV.

Finally, we also considered production of EW particles and investigated the
mono-photon, -$Z$ and -$W$ plus missing momentum signals. We have performed a detailed
analysis for gravitino-pair production, showing  that the ratios of
the different vector bosons in the final state might reveal information about 
left- and right-handed couplings. We have reinterpreted the
mono-photon analysis at $\sqrt{s}=8$ TeV, and found a similar
limit as in the mono-jet analysis in the case where all SUSY particles
except the gravitino are very heavy. For lighter squark masses, the
limits are lower. We have concluded by presenting the outlook for the LHC Run-II.

\section*{Acknowledgments}

This work has been performed in the framework of the ERC grant 291377
``LHCtheory: Theoretical predictions and analyses of LHC physics:
advancing the precision frontier''  and of the FP7 Marie Curie Initial Training Network MCnetITN (PITN-GA-2012-315877).
It is also supported in part by the Belgian Federal Science Policy Office through the Interuniversity Attraction Pole P7/37. 
The work of AM and FM is supported by the IISN ``MadGraph'' convention
4.4511.10 and the IISN ``Fundamental interactions'' convention 4.4517.08.
KM and BO are supported in part by the Strategic Research Program ``High Energy
Physics'' and the Research Council of the Vrije Universiteit Brussel.  

\bibliography{library}
\bibliographystyle{JHEP} 

\end{document}